\documentclass[twocolumn, floatfix, superscriptaddress]{revtex4}

\usepackage{graphicx}
\usepackage{amsmath}
\usepackage{graphicx}
\usepackage{braket}
\usepackage{minitoc}
\usepackage{dirtytalk}
\usepackage[qm]{qcircuit}
\usepackage[colorlinks=true,linkcolor=blue]{hyperref}
\usepackage[capitalise]{cleveref}
\crefname{appsec}{Appendix}{Appendices}
\usepackage[version=3]{mhchem}
\usepackage{mathtools}
\usepackage[titletoc,title,page]{appendix}
\usepackage{algorithm}
\usepackage{algorithmicx}
\usepackage{algpseudocode}

\begin{document}

\date{\today}
\author{Kentaro Yamamoto}
\email{kentaro.yamamoto@cambridgequantum.com}
\affiliation{Cambridge Quantum Computing Ltd., 9a Bridge Street, CB2 1UB Cambridge, United Kingdom}
\author{David Zsolt Manrique}
\affiliation{Cambridge Quantum Computing Ltd., 9a Bridge Street, CB2 1UB Cambridge, United Kingdom}
\author{Irfan Khan}
\affiliation{Cambridge Quantum Computing Ltd., 9a Bridge Street, CB2 1UB Cambridge, United Kingdom}
\author{Hideaki Sawada}
\affiliation{ Advanced Technology Research Laboratories, Nippon Steel Corporation, 20-1 Shintomi Futtsu, Chiba, 293-8511 Japan
}
\author{David Mu\~{n}oz Ramo}
\email{david.munoz.ramo@cambridgequantum.com}
\affiliation{Cambridge Quantum Computing Ltd., 9a Bridge Street, CB2 1UB Cambridge, United Kingdom}

\title{Quantum hardware calculations of periodic systems with partition-measurement symmetry verification: simplified models of hydrogen chain and iron crystals}

\begin{abstract}
    Running quantum algorithms on real hardware
    is essential for understanding their strengths and limitations,
    especially in the noisy intermediate scale quantum (NISQ) era.
    Herein we focus on the practical aspect of quantum computational calculations
    of solid-state crystalline materials
    based on theory developed in our group
    by using real quantum hardware
    with a novel noise mitigation technique referred to as
    partition-measurement symmetry verification,
    which performs post-selection
    of shot counts based on $Z_{2}$ and $U_{1}$ symmetry verification.
    We select two periodic systems with
    different level of complexity for these calculations.
    One of them is the distorted hydrogen chain as
    an example of very simple systems,
    and the other one is iron crystal in the BCC and FCC phases as it is considered to be inaccessible
    by using classical computational wavefunction methods.
    The ground state energies are evaluated
    based on
    the translational quantum subspace expansion (TransQSE) method for
    the hydrogen chain,
    and periodic boundary condition adapted VQE
    for our iron models.
    % In addition to the usual state preparation and measurement noise mitigation,
    % we apply a novel
    % noise mitigation technique, 
    By applying these techniques for the simplest 2 qubit iron model systems, the correlation energies obtained by
    the hardware calculations agree with
    those of the state-vector simulations within $\sim$5 kJ/mol.
    Although the quantum computational resources used for those experiments
    are still limited,
    the techniques applied to obtain our simplified models will
    be applicable in essentially the same manner to more complicated cases as
    quantum hardware matures.
    % the systematic resource reduction applied to obtain our simplified models
    % will give us a way to scale up by rolling approximations back as quantum hardware matures.
\end{abstract}

\maketitle

%==============================================================================
\section{Introduction}
\label{sec:introduction}
%==============================================================================

%\begin{itemize}

%\item In the discussion about the preference of the community for DFT methods, mention that despite the problems, some researchers are working on wavefunction methods for classical PBC calculations that consider electron correlation. Cite papers from Andreas Gruneis (CCSD-PBC \cite{gruneis_cc_pbc2}), Garnet Chan (PySCF \cite{chan_ucc_pbc}), Ali Alavi (FCIQMC \cite{alavi_fci_pbc}) and Cesare Pisani (PBC-MP2, \cite{pisani_mp2_pbc}).

% \item (This maybe in the discussion section). Discuss the issue of needing large amplitudes to get significant results from the IBM experiments

% \item Say That PBC algorithms have to be tested on real hardware to understand strengths and limitations, and this is the purpose of this paper

% \item Provide motivation for the absence of band structures in this paper: we focus here on the challenges of calculating total energies, which are important for a wide range of problems (phase stability, defect formation energies, etc.). Band structures will be studied in future work.

%\end{itemize}

%Since the first quantum chemistry simulation was executed on a quantum computer,\cite{aspuru2005simulated}
``Quantum computational chemistry'' is expected to be
a promising alternative to its conventional (classical) counterpart due to the efficient way in which quantum computers handle large Hilbert spaces.\cite{aspuru2005simulated,mcardle2020quantum}
%\cite{nielsen2002quantum}.
% the quantum computer has a natural way of
% implementing the electronic correlation mapped onto quantum bits (qubits) providing exponentially large computational
% basis with respect to the number of qubits.
% The development of quantum hardware is still in its early stage.
Development in this field has been significantly accelerated by the implementation of cloud-accessible quantum computers from several providers.
Quantum hardware is still in its early stages,
with devices displaying modest numbers of noisy qubits (typically 50--100) that prevent the application of
error correction schemes.
These near-term devices are frequently referred to as Noisy Intermediate Scale Quantum (NISQ) \cite{preskill2018quantum}
devices.

Variational algorithms are believed to be the most suitable techniques for NISQ devices by constructing a hybrid quantum-classical setup,
in which a relatively shallow parameterized quantum circuit performs
heavy tasks such as encoding correlated molecular wavefunctions
to calculate the expectation value of the energy, while the classical computer
collects the data from the quantum computer to optimize the parameters
within the variational loop.
The Variational Quantum Eigensolver (VQE)
algorithm
\cite{peruzzo2014variational}
is one of the most frequently used
variational algorithms to run quantum chemistry simulations on NISQ devices.

In the VQE algorithm,
% as they combine relatively low circuit depths with noise mitigation features.
% In particular, the VQE algorithm consists of an optimization procedure run by a classical
% computer, where the classical computationally heavy tasks
% (e.g., cost function evaluation) are performed with the quantum computer.
we consider a type of cost function $E(\boldsymbol{\theta})$
depending on a set of
parameters $\boldsymbol{\theta}$
describing an associated trial wavefunction $\ket{\Psi(\boldsymbol{\theta})}$,
which is expressed as
\begin{equation}
    E(\boldsymbol{\theta})
    =
    \braket{
        \Psi(\boldsymbol{\theta}) | \hat{H} | \Psi(\boldsymbol{\theta})
    }
    ,
    \label{eq:costf}
\end{equation}
where $\hat{H}$ is a Hamiltonian describing the system.
The trial wavefunction $\ket{\Psi(\boldsymbol{\theta})}$ is actually implemented
into the quantum computer in the following form
\begin{equation}
    \ket{\Psi(\boldsymbol{\theta})}
    =
    \hat{U}(\boldsymbol{\theta}) \ket{\Psi_{0}}
    ,
    \label{eq:ansatz}
\end{equation}
where $\ket{\Psi_{0}}$ corresponds to an initial state that should be
easily prepared.
%In the context of chemistry,
%$\ket{\Phi_{0}}$ is usually set to be the Hartree--Fock state.
The unitary operator $\hat{U}(\boldsymbol{\theta})$ is implemented as
a parameterized quantum circuit,
or ansatz.

While molecular ground state calculations have been focused on 
as the target of early-stage quantum simulations
\cite{mcardle2020quantum,bauer2020quantum},
a number of solid-state quantum calculations have been performed
by employing Periodic Boundary Conditions (PBC)
\cite{cerasoli2020quantum,liu2020simulating,yoshioka2020variational,kais_nn_pbc, nori_nn_pbc,google_pbc,aleiner2020accurately,manrique2020momentum,mizuta2021deep}. These simulations, however, have been performed either on emulators run on classical machines, or on simplified tight-binding models on quantum hardware. Here, we consider the electronic structure hamiltonian explicitly.
Solid state calculations are, in general, more expensive than
molecular cases, because the Hamiltonian describing the system
depends not only on the spin-orbitals but also on the
reciprocal space points (k-points).\cite{evarestov2007quantum}
Most of the efforts devoted to classical computational
PBC calculations have been made based on Density Functional Theory (DFT) \cite{koch2015chemist, martin2020electronic}.
For example, first principles evaluation of the properties of solid-state iron materials has been addressed with DFT methods in many works
\cite{kubler1981magnetic,jiang2003carbon,wang2006finite,domain2004ab,fors2008nature,nguyen2018first}.
Such metallic solid state calculations are known to be
sensitive to the choice
of DFT exchange-correlation functional
without any way to systematically improve the results, in contrast with wavefunction methods where a clear methodology may be followed to obtain increasingly accurate simulations.
Using again the iron example, LDA cannot identify the most stable
phase of this metal and more complex functionals (GGA in this case) are needed to address this problem \cite{asada1992cohesive}.
Despite the computational difficulty in scaling,
wavefunction methods have also been reported
\cite{mcclain2017gaussian,gruneis_cc_pbc2,mcclain2017gaussian,alavi_fci_pbc,pisani_mp2_pbc}
to characterize the effect of electronic correlation,
which is known to be important in, for example,
the interpretation of the behavior
of Mott insulators \cite{mott1937discussion}.
Quantum chemistry calculations on quantum hardware
can potentially extend the
applicability of wavefunction methods
because of the capability of addressing an exponentially growing computational basis
with a relatively small number of qubits.

Even at this early stage,
it is essential to perform real hardware experiments
for quantum computational methods to understand the limitations to be
overcome, because actual quantum and/or classical computational
costs involved in variational algorithms
are not only bound by theoretical computational complexity results
\cite{arora2009computational},
but they may also be influenced by the characteristics of the hardware used.
Such information should be continuously updated to be
utilized for helping hardware development,
so that we can achieve quantum advantage as soon as possible.
Our motivation to write this paper is framed in this context, where we aim to perform benchmark calculations for representative models in order to have a glimpse of the current state of quantum computing techniques for the study of periodic systems. In particular, we set two targets of PBC calculations as follows:
i) distorted hydrogen chain as one of the simplest periodic systems,
and ii) iron crystals as a representative complex system
that will not be easily accessible by classical computational
wavefunction methods.

Quantum hardware development
is in its early stages
and is probably not mature enough to find quantum advantage in the field of quantum chemistry at this moment.
Here instead we
% show the estimate of how much resources is required
%for the accurate description of iron property, and
demonstrate techniques to reduce the quantum
computational resource requirement
by providing a series of (approximate) techniques
to allow us the execution of
quantum hardware calculations with currently available NISQ devices and the study of their performance and challenges in the simulation of periodic systems. The reduced capabilities of current devices force us to use a series of drastic approximations on our models. In order to reduce qubit number requirements and circuit depth, we are restricted to small basis sets, which capture only a fraction of the correlation energy for our systems, and small k-point meshes, which are insufficient to reach the thermodynamic limit.
However, these approximations can be rolled back
as the hardware improves in the future to scale up the complexity of the
target system.

We demonstrate that noise mitigation techniques are
essential for obtaining accurate energies from
NISQ devices.
In particular, we show that State Preparation And Measurement (SPAM) \cite{jackson2015detecting}
noise mitigation improves the accuracy of the energy obtained from quantum hardware, compared to the
exact value obtained by state-vector simulations
on a classical computer. In addition,
we introduce a novel noise mitigation technique called
Partition-Measurement Symmetry Verification (PMSV)
% \todo{reference?}
which significantly improves our results.
As we will demonstrate later, the combination of these noise mitigation techniques
gives us energy values within
$\sim$5 kJ/mol of
agreement with theoretical values in the PBC-adapted VQE calculations for our iron models.
%which is regarded as a reliable starting point
%for the future development of quantum solid-state calculations
%towards quantum advantage.

We focus on the challenges of calculating total energies
in this particular paper,
which are important for a wide range of problems including phase stability or
defect formation energies.
Quantum algorithms for many-body band structure calculations and derived properties will be studied in future work.

% In this paper we perform hardware experiments of
% periodic systems based on the the theory and techniques discussed
% in our previous work.\cite{manrique2020momentum}
% It consists of two hardware-experimental approaches,
% namely, application of the TransQSE method \cite{manrique2020momentum}
% to an H$_{2}$ chain system, and application of PBC-VQE to
% iron crystals.
% The former demonstrates the reduction of resource requirement
% by making use of the translational symmetry of localized orbitals.
% The latter illustrates the systematic resource reduction of
% expensive metallic calculations.
% In both cases, noise mitigation technique is essential to
% obtain the reliable energies.

This paper is organized as follows.
In Sec. \ref{sec:method},
we provide the computational method of the simulations.
In Sec. \ref{sec:application},
we provide our algorithm benchmarking on simple hydrogen lattices,
followed by that of iron crystals
starting with classical calculations
to determine the best way to simplify the system for
calculations using quantum algorithms.
Section \ref{sec:conclusions} concludes this paper with a summary of our findings and an outlook for the future.

%==============================================================================
\section{Methods}
\label{sec:method}
%==============================================================================

%In this section we briefly explain the PBC-adapted NISQ algorithms.
%To this end, we first explain the VQE algorithm in a general context.
%Extensive reviews of quantum computational chemistry including VQE
%are available in References \cite{mcardle2020quantum,bauer2020quantum}
%for example.
%And then we briefly review the PBC-adapted VQE
%and the translational quantum subspace expansion (TransQSE) method.
%Further details are available in our recent paper
%\cite{manrique2020momentum}.

%---------------------------------------------------------------
% \subsection{Computational details}
%\label{sec:methodology}
%---------------------------------------------------------------

In this section, we describe the methodology used to perform our experiments on quantum hardware.
% \todo{explain the choice of algorithms}
In order to run these experiments, we follow a pragmatic line and, for each system considered, we choose a quantum algorithm that allows us to create the simplest quantum circuit compatible with the particular features of the system considered and thus ensures that it will be reliably executed on the quantum computer. Following this criterion, 
Translational Quantum Subspace Expansion (TransQSE)
is applied for the hydrogen chain,
whereas
VQE with unitary coupled cluster singles and doubles with PBC
(UCCSD--PBC) ansatz is used for iron crystals.
The main gain is in using conservation of crystal momentum to simplify the ansatz, by discarding redundant excitations. This reduction suppress the scaling of the number of parameters in ansatz from $O(N^{4}L^{4})$ to $O(N^{4}L^{3})$, where $N$ and $L$ are the number of orbitals in a primitive cell and the number of k points, respectively.
For further details about these
quantum computational theories,
see Ref. \cite{manrique2020momentum}.

This section is structured as follows:
first, we describe the computational details used in common for the distorted hydrogen chain and iron crystals, including Hamiltonian construction, details of the quantum hardware employed and classical optimization schemes chosen. Then, we give details of the noise mitigation techniques applied for this work.
Afterwards, we explain the system-dependent setups and simplifications that guide our choice of quantum algorithm for each model considered. 

\subsection{General Technical Details}
\label{sec:method:technical}

Similarly to the molecular methodology,
the PBC-adapted second-quantized Hamiltonian requires
a set of electronic integrals.
% as shown in Eq.. \eqref{eq:khamiltonian}.
We employ localized atomic Gaussian basis set adapted to translational symmetry \cite{mcclain2017gaussian}.
In contrast to the more popular use of plane wave basis sets
in condensed matter simulations,
it requires significantly smaller number of basis functions
with the expensive integral evaluation costs as a drawback.
Gaussian basis sets are favourable for quantum computations in the NISQ era,
because a smaller number of basis functions leads to a smaller number of qubit
and circuit depth requirements.
We use the Jordan--Wigner scheme \cite{jordan1993Paulische}
to encode a chemistry Hamiltonian
and an ansatz expressed with fermionic operators
to the corresponding qubit operators to be used in
Eq. \eqref{eq:costf} and Eq. \eqref{eq:ansatz}, respectively
\cite{mcardle2020quantum,bauer2020quantum}.

The present calculations are dependent on a number of software packages.
The periodic molecular integral evaluations associated with the
Hartree--Fock (HF) calculations
are performed by using the classical computational chemistry package
\texttt{PySCF} version 1.7.2.\cite{sun2018pyscf}.
\textcolor{black}{Gaussian density fitting is used for efficiently handling electron integrals. To facilitate convergence,
primitive Gaussian functions with
exponents less than 0.1 are discarded
to avoid severe linear dependency. Divergence in exchange energy at the center of the Brillouin zone of PBC-adapted HF or hybrid DFT is addressed by the Ewald correlation scheme.}
The same tool is used to perform
reference  (classical) post-HF calculations
\cite{helgaker2014molecular,szabo2012modern}
such as
k-point dependent 
M{\o}ller--Plesset perturbation theory (MP2)
and coupled cluster singles and doubles (CCSD).
PBC-adapted quantum computational methods
(TransQSE and PBC-adapted VQE) are implemented into
our quantum computational chemistry
software \texttt{InQuanto} \cite{inquanto}, combined
with the \texttt{OpenFermion} package \textcolor{black}{version 0.11.0} \cite{mcclean2020openfermion}
to facilitate fermionic operations.
Quantum circuits are optimized and compiled for each quantum backend with the retargetable quantum compiler \texttt{tket}
\cite{sivarajah2020tket}
via its \texttt{Python} interface, \texttt{pytket} \textcolor{black}{version 0.6.0}.
Quantum simulations running on the classical computer
are performed with
\texttt{Qulacs} \cite{suzuki2020qulacs}
for state-vector simulations and
\texttt{QasmSimulator} of \texttt{Qiskit}
\cite{Qiskit} for shot-based sampling simulations, \textcolor{black}{both via \texttt{pytket} interface}.

%Noisey simulation to emulate real hardware with a noise model is also performed by using \texttt{QasmSimulator},
%which is useful for preliminary simulation of the hardware experiments.
% Hardware experiments are performed on the devices
% made available over the cloud by IBM Quantum.
% We perform noisy simulation by using
% the three different noise model generated from quantum devices identified with
% the names to emulate for noisy simulation.
% \begin{itemize}
%     \item \texttt{ibmq\_5\_yorktown} (5 qubits, 8QV)
%     \item \texttt{ibmq\_rome} (5 qubits, 32QV)
%     \item \texttt{ibmq\_casablanca} (7 qubits, 32QV)
% \end{itemize}

To execute variational optimization, we employ
two classical optimization algorithms:
% PBC-adapted VQE is performed either by using
Rotosolve
\cite{ostaszewski2021structure}
and
Stochastic Gradient Descent (SGD).
% Quantum Natural Gradient (QNG)
\cite{stokes2020quantum,yamamoto2019natural}.
Rotosolve is a gradient-free optimization algorithm based on machine-learning techniques
originally designed for hardware-efficient
ans\"{a}tze,
whereas the SDG algorithm requires gradient evaluated also by
the quantum computer \cite{harrow2019low}.
% whereas the QNG algorithm is regarded as a quantum counterpart of
% the natural gradient method, which requires gradient and
% Fubini-Study metric tensor \cite{harrow2019low}.
Variational optimization of TransQSE is performed by using Rotosolve,
whereas VQE with UCCSD-PBC ansatz is performed by using both Rotosolve and SGD.

Quantum hardware experiments are performed by using
\texttt{ibmq\_casablanca}
% v1.1.14 made available
via cloud service,
which is one of the IBM Quantum Falcon Processors.
The number of shots is chosen to be
$24000$ for each
quantum circuit measurement process.

The approximate correlation energy $E_{\mathrm{corr}}$
defined as the difference between post-HF and HF energies is the primary
target in our quantum hardware experiments.
For convenience, hereafter we use
$\Delta E(\boldsymbol{\theta})$ defined as
\begin{eqnarray}
    \Delta E(\boldsymbol{\theta})
    =
    % \langle \Psi(\boldsymbol{\theta}) |
    % \hat{H} | \Psi(\boldsymbol{\theta}) \rangle
    E_{\mathrm{total}}(\boldsymbol{\theta})
    -
    E_{\textrm{HF}}^{\circ}
    ,
    \label{eq:delta-e}
\end{eqnarray}
where $E_{\textrm{HF}}^{\circ}$ denotes the
HF energy calculated with the classical computer
and
$E_{\textrm{total}}(\boldsymbol{\theta})$
refers to the total energy
calculated by using Eq. (\ref{eq:costf}) with the
chemistry Hamiltonian on quantum hardware
or simulator, which may be influenced by
the noise and/or stochastic error.
If these errors are negligible,
the value of $\Delta E(\boldsymbol{\theta})$ with optimal parameters
$\boldsymbol{\theta}$ is equivalent to
$E_{\mathrm{corr}}$,
which must be a negative value.
However, noise in the NISQ device can make it even a positive value,
because noise-induced high energy excited states may contaminate the calculated ground state wavefunction.
% In some cases one can work around such a positive value by using $E_{\mathrm{total}}(\textbf{0})$ as an
% lternative to $E_{\mathrm{HF}}^{\circ}$ in Eq. (\ref{eq:delta-e})
% as in Ref. \cite{lotstedt2021calculation}.
% This type of heuristic noise mitigation is avoided in this paper to examine the effect of the noise as it is.

\subsection{Noise Mitigation Methods}
\label{sec:method:noise}

% It is essential to perform noise mitigation
% for running quantum algorithms on the NISQ devices.
By the nature of the chemistry-motivated ans\"{a}tze
\cite{mills2020application}
used in this study,
the effect of hardware noise may be understood in terms of involvement of states
that should not participate in the ground state wavefunction, like those not guaranteeing particle number conservation (ionized configurations) or involving a different spin sector (``spin-flipped'' states) than the one corresponding to the ground state.
Noise mitigation techniques with these ans\"{a}tze work to
partially suppress these type of symmetry violations.
% and therefore drive the correlation energy values below the raw counterpart, and towards the negative values one would expect from such a quantity at the $\theta$ amplitude optimum value.

In the present paper we employ two types of
noise mitigation techniques to post-process the
resulting shot counts.
The first technique is
SPAM mitigation implemented
in \texttt{pytket},
whereas the second technique is
PMSV implemented in
\texttt{InQuanto}.
% (\textbf{Details on the noise mitigation should be given})

SPAM mitigation \cite{jackson2015detecting} assumes that noise-induced errors
are independent on the circuit to be executed,
but only occur during the state preparation and measurement steps.
Therefore the density matrix is preliminarily sampled
to get the noise profile of the device.
Then, application of the inverse of the density matrix to the quantum state
would suppress the error caused by these noise channels.

PMSV is a novel technique we have developed to symmetry-verify quantum calculations in an efficient way.
Molecular symmetries can be represented as a linear combination of Pauli strings. 
Symmetries such as mirror planes ($Z_2$) and electron-number conservation ($U_1$) can be represented by a single Pauli string that tracks the parity of the wavefunction \cite{yen2019exact}. One can exploit these symmetries to apply qubit tapering techniques if the system is described by an Abelian point group \cite{bravyi2017tapering, setia2020reducing}. Alternatively, point group symmetry may be applied to mitigate noise. In PMSV, we perform symmetry verification by taking advantage of the commutation of the Pauli symmetries with terms in the Hamiltonian. We can partition these Hamiltonian terms and Pauli symmetries into groups of commuting Pauli strings \cite{gokhale2019minimizing, cowtan2020generic}. If each Pauli string in the partitioned set commutes with the symmetry operator, then each measurement circuit that measures that commuting set can be symmetry-verified without extra quantum resources, discarding the measurements that violate the point group symmetries of the system.
Periodic systems are described by space groups, but by transforming into the reciprocal space, they can be described by point groups, so that we can apply PMSV for solid-state systems as in
molecular systems \cite{manrique2020momentum}. 
A more detailed explanation of this method can be found in Appendix \ref{appendix:sec:pmsv}.

\textcolor{black}{Compared to the other error mitigation methods, PMSV would be characterized as a method exploiting the symmetry with no extra quantum resources, which is scalable and independent of the type of noise. One can perform symmetry verification using mid-circuit measurements, ancilla readouts or the quantum subspace expansion algorithm \cite{Bonet_Monroig_2018}, at the expense of extra quantum resources. Some studies using actual quantum computers \cite{kawashima2021optimizing, google2020hartree} employ McWeeny purification for 2 qubit systems. The one-body reduced density matrix (1-RDM) can be easily purified, but the two-body reduced density matrix (2-RDM) cannot be extended to larger systems without further theoretical investigation. PMSV works with both 1-RDM and 2-RDM in a scalable manner. Other studies use a method relying on the extrapolation technique. For example, this study \cite{gao2021applications} assumes that noise in a basis of states is the same as that in the actual simulation, then applies an inversion matrix to clean the results. PMSV does not have any assumptions on the type of noise.}

%We use \texttt{ibmq\_casablanca}
%(7 qubits, 32QV) for the hardware experiments of VQE,
%and \texttt{ibmq\_bogota}
%(5 qubits, 32QV) for those of QNG optimization,
%where QV denotes the quantum volume
%\cite{cross2019validating}.

%- - - - - - - - - - - - - - - - - - - - - - - - - - - - - - - - 
\subsection{Setup for Hydrogen Molecular Lattices}
\label{sec:method:h2}
%- - - - - - - - - - - - - - - - - - - - - - - - - - - - - - - - 

Let us proceed to the first system to be investigated, a
distorted hydrogen chain.
It is usually taken as the first target of the PBC-adapted variational calculations as in Refs.
\cite{manrique2020momentum,mizuta2021deep,liu2020simulating,yoshioka2020variational, motta_prx_2017}.
More extensive calculations using a quantum simulator running
on the classical computer
for systems including H$_{2}$, He, and LiH lattices
up to three-dimensional lattices
are available in previous work \cite{manrique2020momentum}.

\begin{figure}
    \centering
    \includegraphics[width=0.40\hsize]{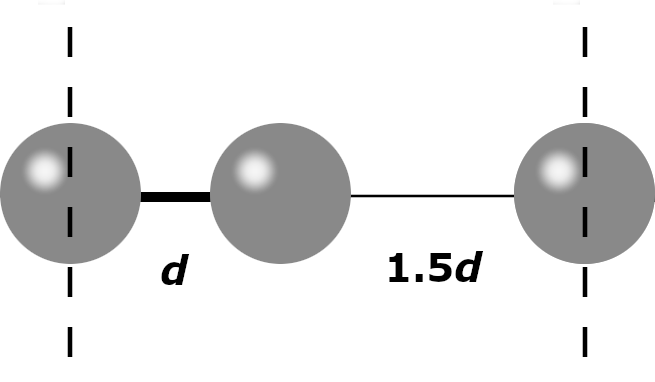}
    \caption{
    Schematic representation of the unit cell
    (indicated by dashed lines)
    of the distorted hydrogen chain with alternating bond lengths,
    $d$ and $1.5d$.
    The bond parameter is set to be $d=0.75$ {\AA} in the present paper.
    }
    \label{fig:h_chain}
\end{figure}

We consider a one-dimensional hydrogen chain with alternating bond length with
$d$ and $1.5d$, with
$d$ being equal to $0.75$ {\AA}
(comparable to the molecular equilibrium H--H distance). With this distortion, we ensure the insulating character of the model.
See Fig. \ref{fig:h_chain}
for the schematic representation of the geometry in the unit cell.

To generate the integrals in the crystalline orbitals,
k-point dependent restricted HF (RHF) with
the STO-3G basis set is used.
The spin multiplicity and the total charge per H$_{2}$ molecule are set
to be 1 (spin-singlet) and 0 (charge-neutral), respectively.
Here we consider a 2 k-point model.
%Energy and amplitude agreements between UCCSD--VQE and CCSD in the state-vector simulations have been confirmed.\cite{manrique2020momentum}
%The VQE results provided in this section are obtained by
%state-vector simulations (classical multiplication of all matrices involved in the calculation), instead of shot-based experiments, with the purpose of obtaining the benchmark results in a faster way.

One could naively consider performing VQE hardware calculations with the UCCSD--PBC ansatz for the 2 k-point model,
requiring
four qubits with six variational parameters to be optimized
after applying the qubit tapering technique based on the
symmetry of the system
\cite{bravyi2017tapering, setia2020reducing}.
However, this setup has been found not
to be feasible on currently available hardware because of the
large circuit depth
of the corresponding ansatz, which includes many excitations with amplitudes of similar absolute value \cite{manrique2020momentum}.

For this reason, instead of VQE with UCCSD--PBC ansatz,
we demonstrate the TransQSE algorithm \cite{manrique2020momentum}
with further approximations for this system.
% In TransQSE, the ground state energy of
% the real-space PBC Hamiltonian
% is computed via a subspace expansion with lattice translational operators.
% The TransQSE method can be understood from two perspectives.
% On one hand, it is a variation of the quantum subspace expansion method
% (QSE) \cite{mcardle2020quantum}.
% On the other hand, it is regarded as a special case of non-orthogonal VQE
% \cite{huggins2020non}.
To perform the TransQSE experiment for the 2 k-point hydrogen chain, the Hamiltonian and the ansatz are transformed to a localized space representation spanned by Wannier orbitals
\cite{wannier1937structure}.
% In this localized representation it has been found that a UCCSD-VQE simulation for the chosen distorted hydrogen chain geometry has two double excitations that are significantly larger than the others.
In this localized representation it has been found that a UCCSD-VQE wavefunction for the chosen distorted hydrogen chain geometry has two double excitations with significantly larger amplitudes than the others.
This is in contrast with the momentum space solution,
where no predominant excitations have been found.
The dominance of these excitations can be also observed in the MP2 approximated amplitudes, and furthermore, the amplitudes of the dominating excitations are the same due to translational symmetry.
% By neglecting all other excitation operators
% excluding these two major ones, 
By neglecting all other excitation operators
apart from these two major ones,
the cluster operator in the UCC ansatz in the real space
$\hat{U}_{\mathrm{R}} = e^{\hat{T} - \hat{T}^{\dagger}}$
is simplified as
%DONE: \todo{Check and add lables to the H2 chain figure; VQE simulation of this in the 1st paper/2nd paper..etc}
\begin{equation}
    \hat{T} \approx \frac{\theta}{2} \left( 
    \hat{a}_{1,1,\uparrow}^{\dagger}\hat{a}_{0,0,\uparrow}  \hat{a}_{1,1,\downarrow}^{\dagger} \hat{a}_{0,0,\downarrow}  + 
    \hat{a}_{0,1,\uparrow}^{\dagger}\hat{a}_{1,0,\uparrow}  \hat{a}_{0,1,\downarrow}^{\dagger} \hat{a}_{1,0,\downarrow} 
   \right ),
    \label{eq:approximated_T_H2R}
\end{equation}
where $\hat{a}_{R,p,\sigma}^{\dagger}$ and $\hat{a}_{R,p,\sigma}$ are the creation and the annihilation operators for the localized orbital $p$ with spin $\sigma$ on the $R$-th H$_2$ molecule in the chain.
Both excitations are inter-molecular and due to the periodic boundary the second term is obtained from the first term by applying a translation operator $\hat{\Lambda}$, which shifts this term to the next unit cell on the right. By taking advantage of the translational operator, instead of the Trotterized UCC ansatz we write the TransQSE ansatz as
\begin{equation}
   \ket{\Psi_{\mathrm{TQSE}}(\theta)}=c_{1} \ket{\Psi_\mathrm{W}(\theta)}+c_{2}  \hat{\Lambda}\ket{\Psi_\mathrm{W}(\theta)}
    \label{eq:tqse_ansatz_H2R},
\end{equation}
where $\ket{\Psi_{\mathrm{W}}(\theta)}=e^{\frac{\theta}{2} 
   \hat{a}_{1,1,\uparrow}^{\dagger}\hat{a}_{0,0,\uparrow}  \hat{a}_{1,1,\downarrow}^{\dagger} \hat{a}_{0,0,\downarrow} - h.c.} \ket{\Phi_{0}}$ and the ground-state energy is found by minimizing the energy function  
\begin{equation}
    E_{\mathrm{TQSE}}(\theta)=\frac{h_{0}(\theta)+h_{1}(\theta)}
   {1 + s_{1}(\theta)}
   ,
    \label{eq:tqse_ground_H2R}
\end{equation}
where 
\begin{eqnarray}
    h_{0}(\theta)&=&\bra{\Psi_{\mathrm{W}}(\theta)} \hat{H} \ket{\Psi_{\mathrm{W}}(\theta)},
    \\
    h_{1}(\theta)&=&\bra{\Psi_{\mathrm{W}}(\theta)}\hat{H} \hat{\Lambda} \ket{\Psi_{\mathrm{W}}(\theta)},
    \\
    s_{1}(\theta)&=&\bra{\Psi_{\mathrm{W}}(\theta)}
   \hat{\Lambda} \ket{\Psi_{\mathrm{W}}(\theta)}
   ,
\end{eqnarray}
as discussed in Ref. \cite{manrique2020momentum}.
Since the excitation only involves a limited range of 4 qubits in $\ket{\Psi_{\mathrm{W}}(\theta)}$, the operators in
Eq. \eqref{eq:tqse_ground_H2R} are reduced to only 4 qubits
from 8 by contraction.
Further resource reduction can be made by exploiting the
symmetry \cite{setia2020reducing}.
This system has a set of symmetry operators $\hat{S}$ as
\begin{equation}
    \hat{S} = 
        \{
        -1\cdot \hat{Z}_{0} \hat{Z}_{2},
        -1\cdot \hat{Z}_{1} \hat{Z}_{3},
        +1\cdot \hat{Z}_{1} \hat{Z}_{2}
        \}
        \label{eq:sym_op}
        .
\end{equation}
The first and second operators are associated with particle conservation for
$\uparrow$ and $\downarrow$ spins, respectively,
and the third one represents the mirror plane spatial symmetry
\cite{yen2019exact}.
We use the first two operators for obtaining the 2-qubit system by tapering off the third and fourth qubits,
and apply the third operator to PMSV noise mitigation.
% Furthermore, 2 more qubits are tapered off from the operators and the ansatz because of the particle conservation for both of $\uparrow$ and $\downarrow$ spins.
After these simplifications are considered, one needs to measure the expectation value of five 2-qubit Pauli strings to determine
$E_{\mathrm{TQSE}}(\theta)$ with the 2-qubit ansatz, 
$
U(\theta)|\Psi_{0}\rangle
=
e^{-i\theta \hat{Y}_{0}\hat{X}_{1}}\ket{0_{0}0_{1}}
$.
To further reduce the number of measurements,
the commuting Pauli strings were measured together using a graph coloring algorithm implemented into \texttt{pytket},
totalling to two circuits to be measured.

To facilitate the variational optimization with hardware,
the term $1 / (1 + s_{1}(\theta))$ in Eq. \eqref{eq:tqse_ground_H2R}
is expanded into a Taylor series as
\begin{eqnarray}
   E_{\mathrm{TQSE}}
   &=&
   \left(
   h_{0}(\theta) + h_{1}(\theta)
    \right)
    \left(
    1 - s_{1}(\theta)
   +O(s_{0}^{2})
    \right)
    .
    \label{eq:tqse_approx}
\end{eqnarray}
As the overlap term $S(\theta)$ is expected to be small
in magnitude due to the relatively low correlation,
taking the first order of
$S(\theta)$ should be a good approximation.
It gives us a steeper function in the large
$|\theta|$ region to accelerate the convergence.
Hereafter we use Eq. \eqref{eq:tqse_approx}
to the first order of $S(\theta)$ as the cost function of
TransQSE.

% DONE: \todo{put here the final numbers from the measurement scripts and the tapered ansatz}

%- - - - - - - - - - - - - - - - - - - - - - - - - - - - - - - - 
\subsection{Setup for Iron Crystals}
%- - - - - - - - - - - - - - - - - - - - - - - - - - - - - - - - 

\begin{figure}
    \centering
    \includegraphics[width=0.80\hsize]{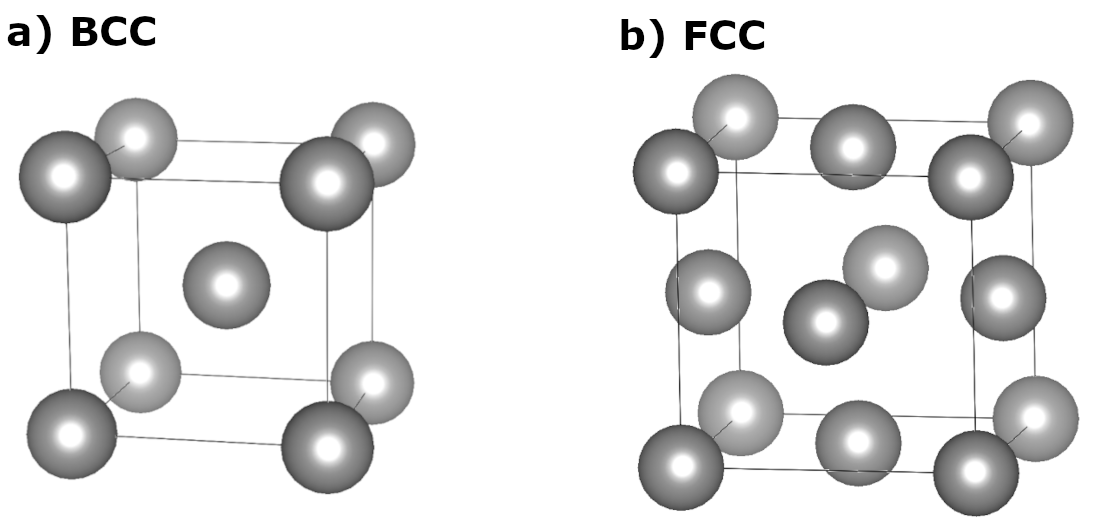}
    \caption{Schematic representation of the unit cell of (a) body-centered cubic
    (BCC) and (b) face-centered cubic (FCC) crystal structures
    \cite{momma2011vesta}.
    }
    \label{fig:bcc_fcc}
\end{figure}

To set a practical starting point of the quantum computational
iron calculations, here we compute the energy difference between
the body-centered cubic [BCC, Fig. \ref{fig:bcc_fcc}(a)]
and face-centered cubic [FCC, Fig. \ref{fig:bcc_fcc}(b)]
crystal structures,
% DONE: \todo{Check the copy right of this figure. Generate them by ourselves if needed}
We focus on the ferromagnetic (FM) phase
of iron in both BCC and FCC lattices,
although it is known that the ground state of the FCC cell
is anti-ferromagnetic (AFM)
\cite{kubler1981magnetic,wang2006finite,nguyen2018first}. This is a fair approximation
because the energy difference between AFM-FCC and FM-FCC
is fairly small
\cite{nguyen2018first}.
Thus, we consider only
the FM-BCC structure with only one possible
spin configuration, which is technically simpler as it allows us to consider a small simulation cell.
Hereafter we refer to this BCC--FCC energy difference without
mentioning their magnetic features.
We shall leave the consideration of
AFM-FCC structures with a variety of
spin configurations for future studies. 
This is due to the increased resources required,
as the model must be constructed with prohibitively
large super-cells,
and single-determinant HF solutions are bad starting points for the calculations.

% We focus on the energy difference between
% the ferromagnetic body-center cubic
% (FM-BCC) and the ferromagnetic face-center cubic % (FM-FCC) structures of
% iron crystal at the optimized lattice constants.

%or unrestricted density functional theory (U-DFT). The
%Perdew, Burke, and Ernzerhof (PBE) exchange-correlation
%functional \cite{perdew1996generalized} is used for U-DFT.

% Excitation operators for PBC-adapted CCSD and UCCSD are classified into
% three types, namely,
% i) excitations between MOs of the $\Gamma$-point,
% ii) excitations between MOs of same k-point,
% and
% iii) excitations betweens MOs of different k-points.
% The latter two types are specific to the PBC.
% Here we consider excitations among all the k-points, but we set small active space
% for MOs to reduce the computational cost.

%However, such metallic systems would require prohibitively many k-points
%for PBC-adapted VQE method with the quantum hardware currently available.
%Therefore, in this project, we investigate a simplified system
%that allows us to perform hardware experiments of PBC-adapted VQE;
%we regard this simplified system as a starting point for the systematic improvement
%of accuracy by increasing the number of k-points and
%molecular orbitals (MOs) in the UCC ans\"{a}tz
%along with the exponential speedup of the quantum computer.\cite{mcardle2020quantum}

We exploit the simplified model system in the following two steps.
i) Classical CCSD calculations are performed for the system with a
relatively large number of k-points (which may be too many for VQE with currently available hardware)
obtained to characterize the effect of electron correlation on the energy difference.
ii) Simplified model systems with the smaller number of k-points and/or active orbitals are designed via a careful analysis of the larger
system investigated in the step (i).
% We will show that such a simplified system can be actually elaborated.
In contrast to the hydrogen chain case,
we identify a few leading excitations in the cluster
operator in momentum space.
This allows us to construct a simplified UCCSD--PBC ansatz for our VQE calculations.

The resulting simplified model system is subjected to the hardware
experiment of VQE with UCCSD--PBC ansatz \cite{manrique2020momentum}.
%The momentum-space second-quantized Hamiltonian
% $\hat{H}_{K}$ and the UCCSD--PBC ansatz $\hat{U}_{K}$
% are encoded to be the qubit operators.
The momentum-space second-quantized
Hamiltonian $\hat{H}_{\mathrm{K}}$
is expressed as \cite{manrique2020momentum}
% \begin{eqnarray}
%     \hat{H}_{\mathrm{K}} &=&
%     \sum'_{\mathbf{k}_{p}\mathbf{k}_{q}}
%     \sum_{PQ} 
%     h^{\mathbf{k}_p P}_{\mathbf{k}_q Q}  
%     \hat{c}^{\dagger}_{\mathbf{k}_{p} P}\hat{c}_{\mathbf{k}_{q} Q} 
%     \nonumber
%     \\
%     &&
%     +
%     \frac{1}{2}
%     \sum'_{\mathbf{k}_{p}\mathbf{k}_{q}\mathbf{k}_{r}\mathbf{k}_{s}}
%     \sum_{\substack{PQRS}}
%     h^{\mathbf{k}_{p}P, \mathbf{k}_{r}R}_{\mathbf{k}_{q}Q, \mathbf{k}_{s}S}
%     \hat{c}_{\mathbf{k}_{p}P}^{\dagger}\hat{c}_{\mathbf{k}_{r}R}^{\dagger}\hat{c}_{\mathbf{k}_{s}S}\hat{c}_{\mathbf{k}_{q}Q}
%     ,
%     \label{eq:khamiltonian}
% \end{eqnarray}
\begin{eqnarray}
    \hat{H}_{\mathrm{K}} &=&
    \sum'_{PQ}
    h^{P}_{Q}  
    \hat{c}^{\dagger}_{P}\hat{c}_{Q}
    +
    \frac{1}{2}
    \sum'_{PQRS}
    h^{PR}_{QS}
    \hat{c}_{P}^{\dagger}\hat{c}_{Q}
    \hat{c}_{R}^{\dagger}\hat{c}_{S}
    ,
    \label{eq:khamiltonian}
\end{eqnarray}
where
$h^{P}_{Q}$
and
$h^{PR}_{QS}$
are the transformed one- and two-electron integrals of the periodic system,
with
$
P, Q, R, S
$ denoting the composite indices of 
k-point in the Brillouin zone $\mathbf{k}$,
spatial orbitals $p$,
and spin $\sigma$.
Here
$\mathbf{k}=\frac{k^{(1)}}{L_1} \mathbf{b}_1 + \frac{k^{(2)}}{L_2} \mathbf{b}_2 + \frac{k^{(3)}}{L_3} \mathbf{b}_3$ with
$\mathbf{b}_1$,$\mathbf{b}_2$ and $\mathbf{b}_3$ denoting
the reciprocal lattice vectors
and $k^{(1)}$, $k^{(2)}$ and $k^{(3)}$ are integers such that $-\frac{L_{\alpha}}{2}<k^{(\alpha)}\leq\frac{L_{\alpha}}{2}$ for $\alpha=1,2,3$
for the [$L_{1}$~$L_{2}$~$L_{3}$] k-point mesh.
Index $p$
labels orbitals in energy-ascending order of each k-point.
The index is actually determined via a mapping function as
$P = q_{\mathrm{K}}(\mathbf{k}, p, \sigma)$
\cite{manrique2020momentum}.
$\hat{c}_{P}^{\dagger}$
($\hat{c}_{P}$)
is the creation (annihilation) operator at
the orbital $P$.
% This operator is related to the
% creation (annihilation) operator
% $\hat{a}_{\mathbf{R}, p, \sigma}^{\dagger}$
% ($\hat{a}_{\mathbf{R}, p, \sigma}$)
% defined in the real space with the lattice vector
% $\mathbf{R}$, orbital $p$, and spin $\sigma$
% by Fourier transform \cite{manrique2020momentum}.
The prime symbol in Eq. \eqref{eq:khamiltonian}
on the sum indicates that it runs only with indices
satisfying crystal momentum conservation \cite{mcclain2017gaussian}.
The cluster operator of the UCCSD--PBC ansatz
$\hat{U}_{\mathrm{K}}=e^{\hat{T}-\hat{T}^{\dagger}}$
is expressed as
\begin{gather}
    \hat{T} = \hat{T}_{1} + \hat{T}_{2}
    ,
    \label{eq:ccop}
    \\
    \hat{T}_{1} = \sum'_{AI}t^{A}_{I}\hat{c}_{A}^{\dagger}\hat{c}_{I}
    ,
    \\
    \hat{T}_{2} =
    \frac{1}{4}\sum'_{ABIJ}
    t^{AB}_{IJ}
    \hat{c}_{A}^{\dagger}
    \hat{c}_{I}
    \hat{c}_{B}^{\dagger}
    \hat{c}_{J}
    ,
\end{gather}
where $t_{I}^{A}$ and $t_{IJ}^{AB}$ are \textcolor{black}{complex-valued coupled cluster amplitudes in general}, with $I,J$ and $A, B$ being the composite indices for occupied and virtual orbitals, respectively.

DFT calculations are performed to support the CCSD calculations
by checking that the properties are not significantly changed
as the number of k-points increases.
See Appendix \ref{appendix:sec:dft} for the details.

We employ the basis set family of
Los Alamos National Lab (LANL) effective core potentials (ECPs).\cite{dunning1977methods}
LANL-ECPs plus double-zeta basis set
(LANL2DZ \cite{hay1985ab})
is used
for all the iron calculations.
% In addition,
% we use LANL-ECPs plus triple-zeta basis (LANL2TZ
% \cite{hay1985ab,roy2008revised}) for DFT % calculations to check the
% sensitivity to the choice of localized atomic % basis set.
% Primitive Gaussian functions with the exponent % less than 0.1 are discarded
% to improve the SCF convergence.
Single-reference calculations are performed with
the k-point dependent
unrestricted HF (UHF) method implemented
in \texttt{PySCF}.
The number of unpaired electrons is set to be
2 per atom to represent the iron FM phases.
Total charge is set to be neutral.

Technically, UHF calculations are not numerically stable.
From our experience,
this feature tends to be prominent in the larger lattice constant.
If the initial guess of UHF is generated at each lattice constant,
the potential energy curve may become discontinuous due to the
different reference configuration.
To obtain a smooth curve for the consistent analysis,
we start our calculation with small lattice constant with
a locally generated initial guess, and then use the converged
crystal orbitals for calculating the next
point in the lattice-constant-ascending order.

%==============================================================================
\section{Results and discussion}
\label{sec:application}
%==============================================================================

%------------------------------------------------------------------------------
\subsection{TransQSE for distorted hydrogen molecular chain with 2 k-points}
\label{sec:h2_transqse}
%------------------------------------------------------------------------------

% \input{h2_tqse_energies}
%-------------------------------------------------------
%------------------------------------------------------------------------------
\subsubsection{Variational optimization with quantum hardware}
%------------------------------------------------------------------------------

% From the noisy simulation results shown above,
% it is expected that the TransQSE variational optimization can
% locate optimum $\theta$ accurately with shifted potential energy
% by $\sim 40$ kJ/mol.
Here we demonstrate
the variational optimization of TransQSE
with the approximate cost function Eq. \eqref{eq:tqse_approx} to the first order
by using Rotosolve with quantum hardware.
%to confirm this prediction.
Initial guess of the amplitude is set to be $\theta = 10^{-5}$
(virtually HF state but with nonzero amplitude to avoid over simplification
of the quantum circuit by \texttt{tket}).

\begin{figure}
    \centering
    \includegraphics[width=0.99\hsize]{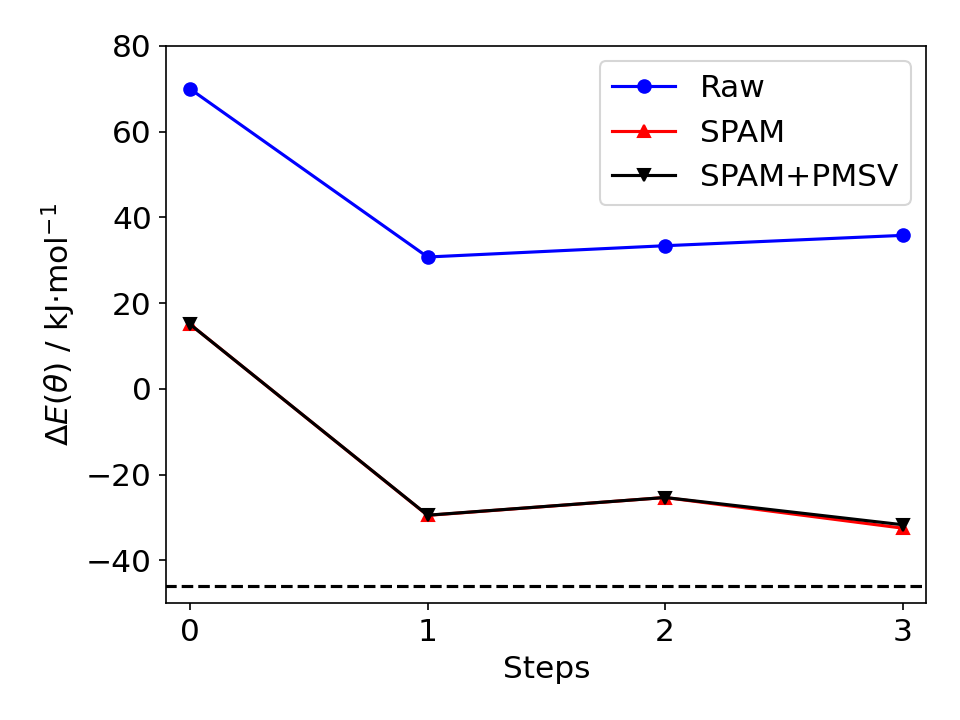}
    \caption{
        Progress of energy estimate
        $\Delta E(\theta)$
        and
        along the variational optimization on hardware
        (\texttt{ibmq\_casablanca}) by using the rotosolve
        algorithm (solid lines with markers).
        $(E_{\mathrm{total}}, E_{\mathrm{HF}}^{\circ})
        % =(-46.0~\mathrm{kJ/mol}, -2676.6~\mathrm{kJ/mol})$
        =(-2.7226\times 10^{3}~\mathrm{kJ/mol}, -2.6766\times 10^{3}~\mathrm{kJ/mol})$
        according to the state-vector simulations.
        The dashed line denotes
        the correlation energy $E_{\mathrm{corr}}$ for the given model Hamiltonian.
        The optimization is performed for the expectation values
        calculated with SPAM- and PMSV-corrected shot counts
        (lower triangles). Those calculated with raw 
        (circles) and SPAM-corrected (upper triangles)
        shot counts are also shown for comparison.
    }
    \label{fig:h2h2_tqse}
\end{figure}

The progress of the energy estimate
is shown as lower triangles (connected by lines as visual guides)
in Fig. \ref{fig:h2h2_tqse}.
At the initial point, $\Delta E(10^{-5}) = 15\pm 5$ kJ$\cdot$mol$^{-1}$,
although it should be virtually 0 kJ/mol by definition.
Hereafter
each uncertainty in the reported expectation values is calculated as the
standard deviation evaluated from the samples for each $\theta$ taken
from the noisy simulator emulating the hardware.
\textcolor{black}{PMSV can influence this uncertainty as some of the measurement outcomes may be discarded, but such an effect was found to be negligible in the present work, as the percentage of discarded shots is $\sim 1$\%
for the H chain.}
The variational experiments converge in three iterations and find the optimal point
$\Delta E(-0.0823) = -31 \pm 4$ kJ/mol,
whose parameter $\theta$ is in good agreement with the state-vector
result $\Delta E(-0.0928)=-46.03$ kJ/mol (the gradient is small $\left.\frac{d(\Delta E)}{d\theta}\right|_{\theta=-0.0928}=13$ kJ/mol), but the energy is again shifted upwards due to the device noise.
The error is larger than those of the iron model systems discussed in Sec. \ref{sec:fesystems}, probably because the energy is evaluated as a product of two expectation values as shown in Eq. \eqref{eq:tqse_approx} to enhance the relative error due to the noise.

% ndicates that even in this simple case the energy estimate
% is significantly influenced by the noise in the quantum device.
Thanks to SPAM correction, the resulting energy is driven to lower values than the
classically evaluated HF energy, that is, $\Delta E < 0$.
The SPAM correction improves the total energy in all the
points in Fig. \ref{fig:h2h2_tqse}
by about 50 kJ/mol on average.
% However, the resulting energy is still
%$~40$ kJ$\cdot$mol$^{-1}$ higher
% than the exact value of state-vector simulation.
PMSV is found to display no significant effect on energy in this particular case.
This is due to the relatively small value of $\theta$,
and SPAM already mitigating the effect of
the symmetry-violating measurement results to be excluded by PMSV.
See Appendix \ref{appendix:sec:pmsv} if one is interested in the details.

As shown above, even in this simple case,
it was found to be difficult to obtain a
quantitative agreement between the energies obtained from state-vector simulations and the quantum device results used in these calculations.
However, the optimal $\theta$ is reproduced accurately within 12\% of
relative error.
Improvements in the agreement between theoretical and experimental $\Delta E(\theta)$ mainly depend on quantum hardware development,
although advances in noise mitigation techniques will also have an impact on these results.
In contrast to the PBC-adapted VQE,
TransQSE can still be compact even with more k-points
by truncating spatial long-range correlations.
The state-vector simulation of TransQSE demonstrates that good approximations of the ground state
energy value are obtained ($\Delta E = -46.03$ kJ/mol) 
with a smaller number of parameters than the analogous PBC-adapted VQE calculation
($\Delta E = -57.73$ kJ/mol \cite{manrique2020momentum}).
However, exactly predicting in which cases
TransQSE is a more feasible choice is generally not trivial. In this sense, cheap pre-processing techniques to screen important angles in the ansatz may help to solve this question.
%It would be interesting to compare the ground state obtained by both TransQSE and PBC-adapted VQE not only by the state-vector simulation but also on quantum hardware.
% We may need to find a good balance of
% many things including circuit depth and 

%------------------------------------------------------------------------------
\subsection{Simulations of iron systems}
\label{sec:fesystems}
%------------------------------------------------------------------------------

%------------------------------------------------------------------------------
\subsubsection{Classical CCSD calculation with a [4~4~4] k-point mesh}
\label{sec:fe_simulations:classical}
%------------------------------------------------------------------------------

%It is not trivial how the electron correlation affects the property
%calculated with the single-reference methods.
First we investigate the effect of electronic correlation on the
BCC--FCC energy difference by performing classical UHF/CCSD calculations.
We will consider a model for these iron phases simplified in such a way that the most important
features in these relatively complex systems
are captured with as much fidelity as possible.

Within the available computational resources, we consider a
UHF/CCSD setup with a [4~4~4] Monkhorst-Pack mesh, with
the active space consisting
of the crystal orbitals with orbital and spin indices
$(p, \sigma) = \{
(8, \uparrow),
(9, \uparrow), 
(6, \downarrow),
(7, \downarrow)
\}$
for each $\mathbf{k}$ value.
These crystal orbitals correspond to
linear combinations of iron $d$ orbitals in the region around the Fermi level.
%one unoccupied and one occupied spin orbital for each spin and k-point pair)
The total number of active electrons and crystal orbitals
are equal to 128 and 256, respectively.
The reference electronic configuration is not uniform between k-points,
but each $\mathbf{k}$ sector contains 0 to 4 electrons in it.
We have indirectly confirmed that this $[4~4~4]$ k-point mesh
is sufficient by using DFT with the same basis set,
because the potential energy curves were not significantly changed
as the number of k-points increased
(see Appendix \ref{appendix:sec:dft}).
The number of k-points is still significantly smaller than the converged values in literature
\cite{jiang2003carbon,nguyen2018first, alnemrat_jpcm_2014, godwal_epsl_2015},
% but it is large enough as the k-point sampling error is small for the purposes of this study.
but it is large enough for the k-point sampling error to be sufficiently small for the purposes of this study.
%This problem will be revisited once we become able to address bigger model systems.

\begin{figure}
    \centering
    \includegraphics[width=0.99\hsize]{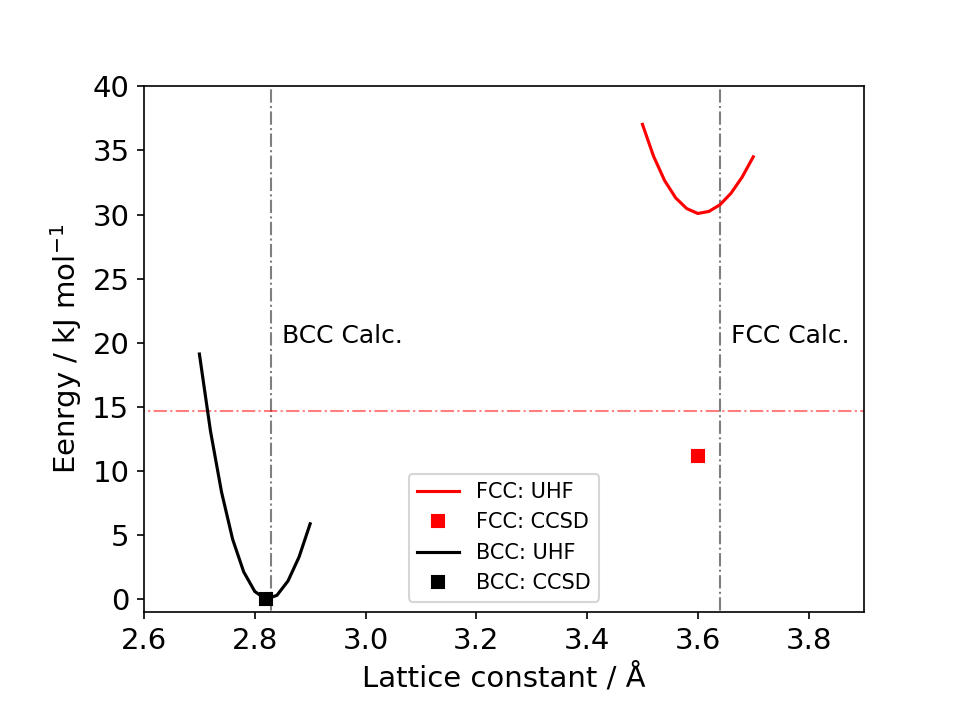}
    \caption{
        Potential energy curves of FM-BCC
        and FM-FCC iron as functions of 
        lattice constants for [4~4~4] k-point mesh
        at UHF level. Also represented are
        single-point calculations at CCSD level
        at optimized lattice constant values
        for BCC and FCC.
        The energy baseline is set so that the lowest energy
        is equal to zero for each UHF and UHF/CCSD.
        The optimized lattice constants
        from the accurate plane-wave basis DFT calculations
        \cite{nguyen2018first}
        are indicated by vertical lines (BCC: 2.83 {\AA}, FCC: 3.64 {\AA}),
        while the BCC--FCC energy difference is represented by the
        horizontal red line (14.67 kJ/mol).
    }
    \label{fig:pes_uhf_ccsd_64k}
\end{figure}

One of the remarkable features in the comparison of UHF vs. CCSD results
is that UHF significantly overestimates the BCC--FCC energy difference,
but CCSD corrects this energy difference
to a value comparable to
accurate plane-wave basis DFT results \cite{nguyen2018first},
which are used as reference values. This trend is illustrated in our calculations of
the potential energy curve as a function of the lattice constant
with UHF for each of FCC and BCC, as shown in Fig. \ref{fig:pes_uhf_ccsd_64k}.
The UHF calculations accurately reproduce the reference lattice constants,
but the estimated BCC--FCC difference ($30$ kJ/mol) 
is significantly overestimated, to almost twice the reference value
as shown in Fig. \ref{fig:pes_uhf_ccsd_64k}.
Such a disagreement is corrected by taking the electronic correlation
into account with CCSD.
The total energy with CCSD/UHF at the optimized lattice constant with UHF
for each of BCC and FCC is shown in Fig. \ref{fig:pes_uhf_ccsd_64k}.
We find that the energy difference
is corrected to $12$ kJ/mol,
which is indeed comparable to the reference value indicated by the
horizontal line in Fig. \ref{fig:pes_uhf_ccsd_64k}.

These results suggest that the FCC phase is more influenced by the electronic correlation than the BCC phase.
% We admit that our iron calculations have a room for improvement.
% The ferromagnetic cases are focused on in this paper,
% The most stable FCC phase is known to be
% antiferromagnetic (AFM-FCC) \cite{nguyen2018first}.
% because of the strong electronic correlation.
We will not provide a detailed interpretation,
% of the BCC--FCC
% energy difference correction by CCSD (Fig. \ref{fig:pes_uhf_ccsd_64k}),
but many competing phases may contribute to increasing
the electronic correlation in the FCC case, in contrast to
the isolated BCC phase \cite{jiang2003carbon}.
Hereafter we focus on this feature to
construct simplified model systems to set a realistic starting point
for experiments on currently available quantum computers.
In other words, our simplifications should take into account that the correlation energy in our simplified FCC model will be larger than that of the BCC model.
% \todo{put the contents in the middle of the first section}
% Estimation of the energy difference between FM-BCC and AFM-FCC(s) would be
% even more sensitive to the choice of DFT functionals.

%Further characterization of the MOs involved in the electron correlation
%will be considered in the future work.

%------------------------------------------------------------------------------
\subsubsection{Simplified system with [2~1~1] k-point mesh}
\label{sec:fesystems_2k}
%------------------------------------------------------------------------------

% It is not yet practical for quantum hardware currently available
% to execute such a large calculation
% as PBC-adapted VQE with the $[4~4~4]$ k-point mesh
% to perform the BCC--FCC energy evaluation in good agreement
% with that of accurate DFT \cite{nguyen2018first}.
% Recall that 256 crystal orbitals 
% [(64 k-points) $\times$ (2 localized MOs) $\times$
% (2 spins)] are required for the case as
% we have discussed above in
% Sec. \ref{sec:fe_simulations:classical}.
%Therefore at present it is not really possible to obtain the accurate
% properties of iron crystals by the quantum computer available today.

As a first step
towards the quantum hardware experiments,
we start from a simplified model system
to adjust the resources needed for PBC-adapted VQE calculations to the constraints imposed by the device we are going to target.
%we extract the simplest system that has
%the feature discussed above, that is,
%single-reference calculation overestimates the
%BCC-FCC energy difference,
%but post-HF calculations would correct it to some extent
%by obtaining larger correlation energy in FM-FCC than FM-BCC.
Let us employ the simplest possible [2~1~1] k-point mesh
to test if this simplified system can reproduce the FCC--BCC features mentioned in Sec. \ref{sec:fe_simulations:classical}.

\begin{figure}
    \centering
    \includegraphics[width=0.99\hsize]{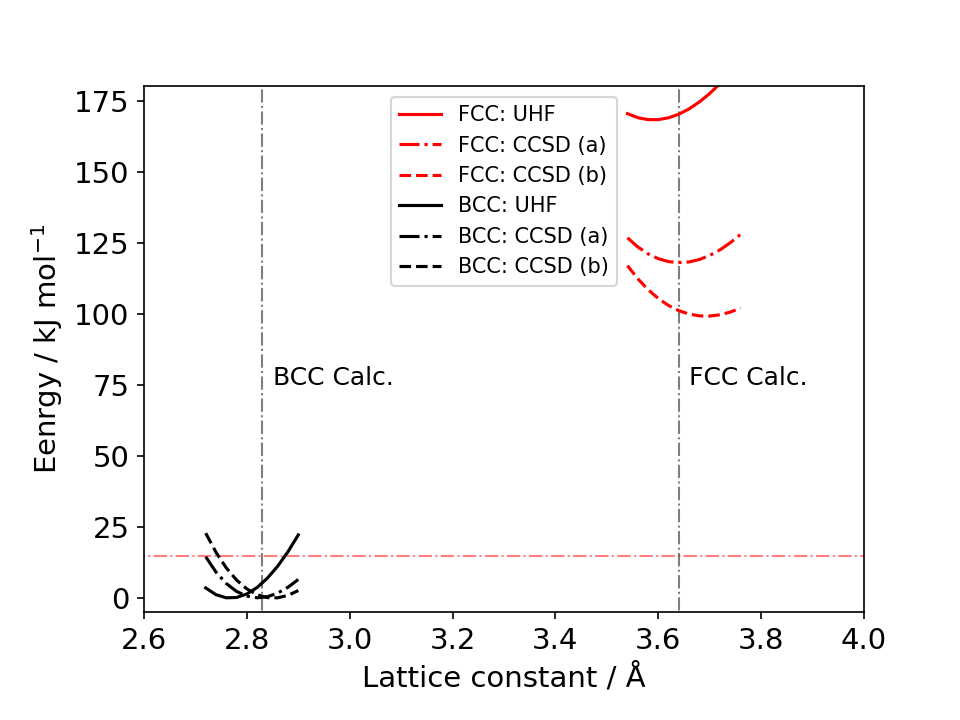}
    \caption{
    Potential energy curves as functions of 
    lattice constants for $[2~1~1]$ k-point mesh
    with UHF, 
    CCSD/UHF with active space [indicated by CCSD(a),
    See Fig. \ref{fig:iron_mo_scheme}],
    and CCSD/UHF with all the valence orbitals
    [indicated by CCSD(b)].
    UHF overestimates the BCC--FCC energy difference (168 kJ/mol),
    and CCSD corrects it to some extent
    [118 kJ/mol for CCSD(a) and 100 kJ/mol for CCSD(b)],
    which is qualitatively the same feature found in the
    accurate case with $[4~4~4]$ k-points.
    The meanings of the vertical and horizontal lines in each panel
    are the same as those in Fig. \ref{fig:pes_uhf_ccsd_64k}.
    }
    \label{fig:pes_uhf_ccsd_02k}
\end{figure}

The potential energy curves calculated with UHF on this model
[Fig. \ref{fig:pes_uhf_ccsd_02k}] indicate
that UHF overestimates the BCC--FCC energy difference,
which is the same trend observed in the [4~4~4] k-point mesh cases.
Although these simplified models are
quantitatively insufficient,
the UHF calculation returns a good estimate of the equilibrium
lattice constant. This implies that
this drastically simplified model is able to qualitatively reproduce the properties of the iron phases in which we are interested.

%The overestimated energy is almost ten times larger than
%the reference value, but it has qualitatively the same tendency.
The potential energy curves calculated with UHF/CCSD, including all the excitation operators in the cluster operator $\hat{T}$, are shown in
Fig. \ref{fig:pes_uhf_ccsd_02k}.
Although the BCC--FCC energy difference is still overestimated,
it is indeed corrected in the right direction.
As shown in Fig. \ref{fig:pes_uhf_ccsd_02k},
similar results are obtained by using the same active space
used for the [4~4~4] k-point mesh case, that is,
$(p, \sigma) = \{
(8,\uparrow), 
(9,\uparrow), 
(6,\downarrow), 
(7,\downarrow)
\}$
for each $\mathbf{k}$.
The reference configuration is given as
\{1, 1, 1, 1\} for $k = 0$ and \{0, 0, 0, 0\} for $k = 1$,
where $k \equiv \mathbf{k} = \frac{k}{2}\mathbf{b}_{1}$.
This suggests that the correlation energy is dominated by
the frontier orbitals as in the molecular cases \cite{fukui1952molecular}. 
Therefore, this simple model is validated and we will employ the [2~1~1] k-point mesh
with the active space for quantum hardware experiments.
Hereafter the optimized lattice constants
(2.84 {\AA} for BCC and 3.64 {\AA} for FCC)
obtained by UHF/CCSD with the active space are used.
The resulting approximate correlation energy
$E_{\mathrm{corr}}$ for each of BCC and FCC is shown
in Tab. \ref{tab:iron_energies}, which is used for comparing with
the quantum computational methods.
\begin{table}
    \centering
    \caption{
        Approximate correlation energy $E_{\text{corr}}$
        of iron crystal calculations with 2 k-points
        obtained by using either CASCI, CCSD or VQE with the UCCSD-PBC ansatz. The active space used is identified:
        \{$(k, 8,\uparrow)$, $(k, 9,\uparrow)$,
        $(k, 6,\downarrow)$, $(k, 7,\downarrow)$\} for $k = 0, 1$
        (see Fig. \ref{fig:iron_mo_scheme}).
        %The comparison between CCSD and VQE
        %with the active space type (2) is made to confirm that
        %they are in good agreement with each other
        %if the same active space is given.
    }
    \label{tab:iron_energies}
    \begin{tabular}{c|c|rr}
        \hline
        Method & excitation operator(s) & \multicolumn{2}{c}{$E_{\text{corr}}$ / kJ$\cdot$mol$^{-1}$} \\
        \hline
        & & \multicolumn{1}{c}{BCC} & \multicolumn{1}{c}{FCC} \\
        \hline
        % CCSD & (1) & -585.64 & -666.19 \\
        CASCI & all & -236.4 & -281.1 \\
        CCSD  & all & -235.2 & -281.1 \\
        VQE   & all & -236.0 & -281.1 \\
        VQE & one$^{a)}$ & -220.5 & -271.5 \\
        \hline
    \end{tabular}
    \\
     $a)$ $
        \hat{c}_{1,8,\uparrow}^{\dagger}
        \hat{c}_{0,9,\uparrow}
        \hat{c}_{1,6,\downarrow}^{\dagger}
        \hat{c}_{0,7,\downarrow}
    $
\end{table}

% The MO energies and the occupation numbers at the optimized lattice constants
% are listed in
% Table \ref{tab:iron_mo_bcc} and
% Table \ref{tab:iron_mo_fcc}
% for BCC and FCC structures, respectively.
% \input{iron_mo_bcc}
 %\input{iron_mo_fcc}

It is required to further simplify the model system,
because the quantum circuits corresponding to
the PBC-adapted VQE with the UCCSD ansatz are
still too deep for the currently available quantum hardware.
To demonstrate the quantum hardware calculations,
let us perform further simplification of the system based on the same
idea applied for the TransQSE calculations, that is,
focusing on the excitation operators with prominent amplitudes.
We performed state-vector simulations of VQE
for iron with the active space mentioned above.
The resulting energies of both BCC and FCC are in good agreement with
the CCSD counterpart (see Tab. \ref{tab:iron_energies}).
\textcolor{black}{The complete active space configuration interaction (CASCI) via exact diagonalization is also performed for comparison.}
The nonzero cluster amplitudes of VQE
are listed in
Tab. \ref{tab:iron_vqe_bcc}
and
Tab. \ref{tab:iron_vqe_fcc}
for BCC and FCC, respectively.
\textcolor{black}{Note that all the amplitudes in the ansatz are of real number as a consequence of the reduction to 2 k-points}.
In both cases, we can find one prominent cluster amplitude
corresponding to a inter-k-point double excitation.
This inter-k-point double-excitation has more than ten times
larger amplitude in magnitude compared to the others.
Therefore
we may define an even more compact active space consisting of 
these four crystal orbitals with an ansatz including one double excitation.
\begin{table}
    \small
    \centering
    \caption{
        Non-zero cluster amplitudes
        $t^{AB}_{IJ}$
        of the PBC-adapted VQE calculation with the
        UCCSD ansatz for
        the 2 k-point model of iron BCC structure with the active space (see text).
    }
    \label{tab:iron_vqe_bcc}
    \begin{tabular}{cccc|r}
    \hline
    $A$ & $I$ &
    $B$ & $J$ &
    \multicolumn{1}{c}{$t^{AB}_{IJ}$} \\
    \hline
    (1, 7, $\downarrow$) &  (0, 7, $\downarrow$) &    (1, 6, $\downarrow$) &  (0, 6, $\downarrow$) &   -0.0060 \\
    (1, 8, $\uparrow$)   &  (0, 8, $\uparrow$)   &    (1, 6, $\downarrow$) &  (0, 6, $\downarrow$) &    0.0160 \\
    (1, 8, $\uparrow$)   &  (0, 8, $\uparrow$)   &    (1, 7, $\downarrow$) &  (0, 6, $\downarrow$) &   -0.0001 \\
    (1, 9, $\uparrow$)   &  (0, 8, $\uparrow$)   &    (1, 6, $\downarrow$) &  (0, 6, $\downarrow$) &    0.0001 \\
    (1, 9, $\uparrow$)   &  (0, 8, $\uparrow$)   &    (1, 7, $\downarrow$) &  (0, 6, $\downarrow$) &   -0.0528 \\
    (1, 8, $\uparrow$)   &  (0, 8, $\uparrow$)   &    (1, 6, $\downarrow$) &  (0, 7, $\downarrow$) &    0.0004 \\
    (1, 8, $\uparrow$)   &  (0, 8, $\uparrow$)   &    (1, 7, $\downarrow$) &  (0, 7, $\downarrow$) &    0.0206 \\
    (1, 9, $\uparrow$)   &  (0, 8, $\uparrow$)   &    (1, 6, $\downarrow$) &  (0, 7, $\downarrow$) &   -0.0709 \\
    (1, 9, $\uparrow$)   &  (0, 8, $\uparrow$)   &    (1, 7, $\downarrow$) &  (0, 7, $\downarrow$) &   -0.0003 \\
    (1, 8, $\uparrow$)   &  (0, 9, $\uparrow$)   &    (1, 6, $\downarrow$) &  (0, 6, $\downarrow$) &    0.0008 \\
    (1, 8, $\uparrow$)   &  (0, 9, $\uparrow$)   &    (1, 7, $\downarrow$) &  (0, 6, $\downarrow$) &   -0.0420 \\
    (1, 9, $\uparrow$)   &  (0, 9, $\uparrow$)   &    (1, 6, $\downarrow$) &  (0, 6, $\downarrow$) &    0.0311 \\
    (1, 9, $\uparrow$)   &  (0, 9, $\uparrow$)   &    (1, 7, $\downarrow$) &  (0, 6, $\downarrow$) &   -0.0002 \\
    (1, 8, $\uparrow$)   &  (0, 9, $\uparrow$)   &    (1, 6, $\downarrow$) &  (0, 7, $\downarrow$) &   -0.4892 \\
    (1, 8, $\uparrow$)   &  (0, 9, $\uparrow$)   &    (1, 7, $\downarrow$) &  (0, 7, $\downarrow$) &   -0.0002 \\
    (1, 9, $\uparrow$)   &  (0, 9, $\uparrow$)   &    (1, 6, $\downarrow$) &  (0, 7, $\downarrow$) &   -0.0003 \\
    (1, 9, $\uparrow$)   &  (0, 9, $\uparrow$)   &    (1, 7, $\downarrow$) &  (0, 7, $\downarrow$) &    0.0058 \\
    (1, 9, $\uparrow$)   &  (0, 9, $\uparrow$)   &    (1, 8, $\uparrow$)   &  (0, 8, $\uparrow$)   &   -0.0630 \\
    \hline
\end{tabular}

\end{table}
\begin{table}
    \small
    \centering
    \caption{
        Non-zero cluster amplitudes
        $t^{AB}_{IJ}$
        of the PBC-adapted VQE calculation with the
        UCCSD ansatz for
        the 2 k-point model of iron FCC structure with the active space (see text).
    }
    \label{tab:iron_vqe_fcc}
    \begin{tabular}{cccc|r}
        \hline
    $A$ & $I$ &
    $B$ & $J$ &
    \multicolumn{1}{c}{$t^{AB}_{IJ}$} \\
        \hline
        (1, 7, $\downarrow$) &  (0, 7, $\downarrow$) &  (1, 6, $\downarrow$) & (0, 6, $\downarrow$) &   -0.0088 \\
        (1, 8, $\uparrow$)   &  (0, 8, $\uparrow$)   &  (1, 6, $\downarrow$) & (0, 6, $\downarrow$) &   -0.0121 \\
        (1, 8, $\uparrow$)   &  (0, 8, $\uparrow$)   &  (1, 7, $\downarrow$) & (0, 6, $\downarrow$) &   -0.0047 \\
        (1, 9, $\uparrow$)   &  (0, 8, $\uparrow$)   &  (1, 6, $\downarrow$) & (0, 6, $\downarrow$) &   -0.0023 \\
        (1, 9, $\uparrow$)   &  (0, 8, $\uparrow$)   &  (1, 7, $\downarrow$) & (0, 6, $\downarrow$) &    0.0040 \\
        (1, 8, $\uparrow$)   &  (0, 8, $\uparrow$)   &  (1, 6, $\downarrow$) & (0, 7, $\downarrow$) &   -0.0113 \\
        (1, 8, $\uparrow$)   &  (0, 8, $\uparrow$)   &  (1, 7, $\downarrow$) & (0, 7, $\downarrow$) &    0.0221 \\
        (1, 9, $\uparrow$)   &  (0, 8, $\uparrow$)   &  (1, 6, $\downarrow$) & (0, 7, $\downarrow$) &   -0.0445 \\
        (1, 9, $\uparrow$)   &  (0, 8, $\uparrow$)   &  (1, 7, $\downarrow$) & (0, 7, $\downarrow$) &   -0.0217 \\
        (1, 8, $\uparrow$)   &  (0, 9, $\uparrow$)   &  (1, 6, $\downarrow$) & (0, 6, $\downarrow$) &    0.0069 \\
        (1, 8, $\uparrow$)   &  (0, 9, $\uparrow$)   &  (1, 7, $\downarrow$) & (0, 6, $\downarrow$) &    0.0050 \\
        (1, 9, $\uparrow$)   &  (0, 9, $\uparrow$)   &  (1, 6, $\downarrow$) & (0, 6, $\downarrow$) &   -0.0320 \\
        (1, 9, $\uparrow$)   &  (0, 9, $\uparrow$)   &  (1, 7, $\downarrow$) & (0, 6, $\downarrow$) &   -0.0083 \\
        (1, 8, $\uparrow$)   &  (0, 9, $\uparrow$)   &  (1, 6, $\downarrow$) & (0, 7, $\downarrow$) &   -0.5839 \\
        (1, 8, $\uparrow$)   &  (0, 9, $\uparrow$)   &  (1, 7, $\downarrow$) & (0, 7, $\downarrow$) &   -0.1017 \\
        (1, 9, $\uparrow$)   &  (0, 9, $\uparrow$)   &  (1, 6, $\downarrow$) & (0, 7, $\downarrow$) &   -0.0053 \\
        (1, 9, $\uparrow$)   &  (0, 9, $\uparrow$)   &  (1, 7, $\downarrow$) & (0, 7, $\downarrow$) &    0.0009 \\
        (1, 9, $\uparrow$)   &  (0, 9, $\uparrow$)   &  (1, 8, $\uparrow$)   & (0, 8, $\uparrow$)   &   -0.0499 \\
        \hline
    \end{tabular}

\end{table}

\begin{figure}
    \centering
    \includegraphics[width=0.80\hsize]{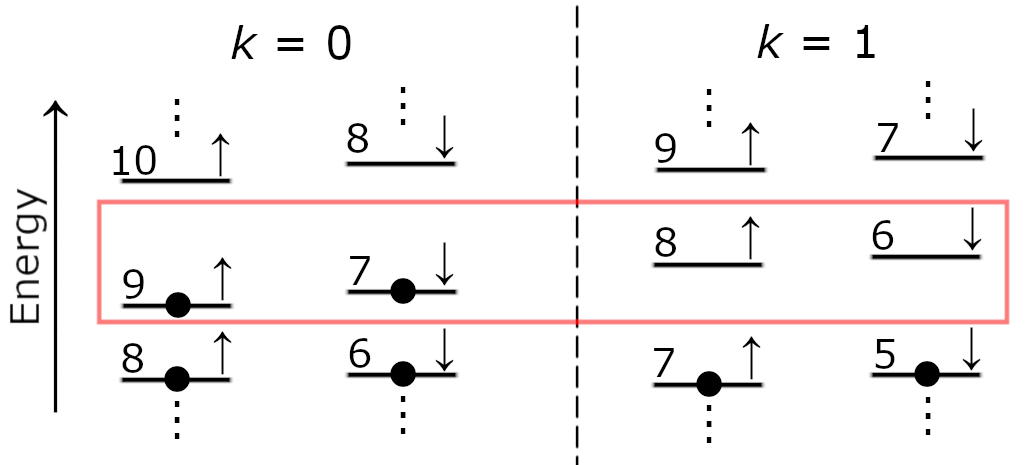}
    \caption{
        Schematic representation of the orbital energy levels
        for the simplified 2 k-point iron model.
        Each horizontal bar represents the energy level with the number corresponding to the orbital index $p$ with spin
        $\sigma = \{\uparrow, \downarrow\}$.
        Occupied orbitals are indicated by the points
        ``$\bullet$''.
        The active space used in the present hardware calculations
        is represented by the red rectangle in color.
    }
    \label{fig:iron_mo_scheme}
\end{figure}

As shown in Tab. \ref{tab:iron_energies},
more than 90\% of the reference correlation energy is reproduced with this
one double excitation operator in both BCC and FCC cases.
More excitation operators in the ansatz can in principle improve the correlation energy,
but such a small difference may not be resolved with the deepened quantum circuit on the device currently available.
Therefore the cluster operator is approximately expressed as
\begin{eqnarray}
    \hat{T}(\theta)
    &\approx&
    \theta
    \hat{c}_{1,8,\uparrow}^{\dagger}
    \hat{c}_{0,9,\uparrow}
    \hat{c}_{1,6,\downarrow}^{\dagger}
    \hat{c}_{0,7,\downarrow}
    ,
\end{eqnarray}
As a consequence,
we have elaborated a simple computational setup
consisting of four active crystal orbitals with one
double-excitation operator of the UCCSD--PBC ansatz with two electrons.
See Fig. \ref{fig:iron_mo_scheme} for the schematic representation of the orbital energy levels and the active space.
The same symmetry operators as in the model system for TransQSE
(Eq. \eqref{eq:sym_op}) are used for
reducing the sytem to an equivalent 2-qubit system
by applying the qubit tapering technique,
and for PMSV noise mitigation.
To determine the expectation value
$\langle \Psi(\theta) | \hat{H}_{\mathrm{K}} | \Psi(\theta) \rangle$,
one needs to measure the expectation value of four 2-qubit Pauli strings
with the ansatz 
$U(\theta)|\Psi_{0}\rangle
= e^{-i\theta \hat{Y}_{0}\hat{X}_{1}}\ket{0_{0}0_{1}}$.
Applying the same partitioning by using \texttt{pytket},
two circuits are actually measured in both BCC and FCC models.
% Note that the number of circuits to be measured
% is less than that of TransQSE case,
% which would be one of the reasons why the accuracy of these systems is different from each other,
%  despite the same number of qubits.
%------------------------------------------------------------------------------
\subsubsection{Variational quantum hardware experiments}
%------------------------------------------------------------------------------

Let us proceed to demonstrate the variational optimization
of the energy by using real quantum hardware
provided through cloud by IBM Quantum.
We need at least $\sim$10 kJ/mol
of accuracy to reliably discuss the BCC--FCC energy
difference, but reaching this threshold has been found to be unfeasible without noise mitigation
in our preliminary calculations, even for this simplified 2-qubit model system.
To realize this accuracy on the NISQ device,
we apply SPAM and PMSV noise mitigation.
By using these techniques for \texttt{ibmq\_casablanca}, a
sufficiently accurate expectation value is estimated within
$\sim 5$ kJ/mol of error compared to the state-vector simulation results.

\begin{figure*}
    \centering
    \includegraphics[width=0.84\hsize]{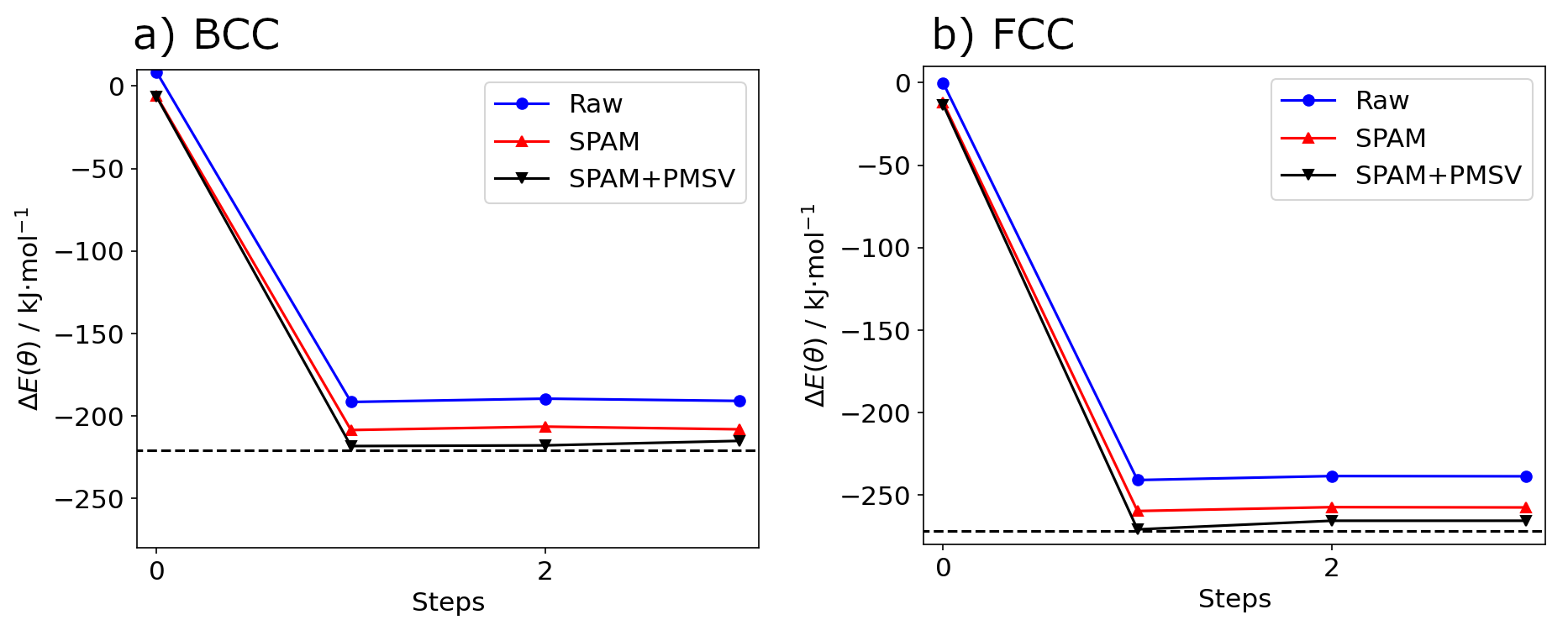}
    \caption{
        Evolution of the $\Delta E$ with respect to the number of steps used by the Rotosolve optimization process.
        $(E_{\mathrm{total}}, E_{\mathrm{HF}}^{\circ})
        % =(-220.5~\mathrm{kJ/mol}, -322636.8~\mathrm{kJ/mol})$
         =(-3.228573\times 10^{5}~\mathrm{kJ/mol}, -3.226368\times 10^{5}~\mathrm{kJ/mol})$
        for BCC and
        $(E_{\mathrm{total}}, E_{\mathrm{HF}}^{\circ})
        %=(-271.5~\mathrm{kJ/mol}, -322473.5~\mathrm{kJ/mol})$
        =(-3.227450\times 10^{5}~\mathrm{kJ/mol}, -3.224735\times 10^{5}~\mathrm{kJ/mol})$
        for FCC according to the state-vector simulations.
        Black points show hardware results obtained with the \texttt{ibmq\_casablanca} device. The line is included as a visual guide to show the
        progress of the correlation energy estimate along the variational optimization.
        The dashed line is the correlation energy $E_{\mathrm{corr}}$ for the given model Hamiltonian.
        FCC result has 51 kJ/mol more correlation energy in magnitude to correct the BCC--FCC energy difference from 168 kJ/mol to 117 kJ/mol, which is qualitatively the same feature found in the accurate case with
        [4~4~4] k-point mesh.
        SPAM and PMSV
        are applied at the same time.
        Raw and SPAM-corrected $\Delta E$ are given for each optimization step
        to show the effect of noise mitigation.
    }
    \label{fig:exp_rotosolve}
\end{figure*}

First we show the results of variational optimization
with Rotosolve.
The progress of correlation energy along the variational optimization
by using Rotosolve is shown in
Fig. \ref{fig:exp_rotosolve}(a)
and
Fig. \ref{fig:exp_rotosolve}(b)
for BCC and FCC, respectively.
In both cases, optimization rapidly converges to the point
that it is in very good agreement with the energy obtained by the
state-vector simulation.
\textcolor{black}{The resulting parameter and the energy are represented as
$\Delta E(-0.5561) = -215 \pm 3$ kJ/mol for BCC and
$\Delta E(-0.6522) = -265 \pm 2$ kJ/mol for FCC.}
The small number of optimization steps owes to the
properties of Rotosolve, which is designed for
systems described by quantum circuits with parameterized rotation gates.
% The uncertainty is estimated to be $\sim 3$ kJ/mol as a standard deviation, thus we can
reliably tell the difference in correlation energy (not the total energy)
between FM-BCC and FM-FCC (51 kJ/mol),
in which FM-FCC has more correlation energy to correct the UHF value indeed
to the right direction.

\begin{figure*}
    \centering
    \includegraphics[width=0.84\hsize]{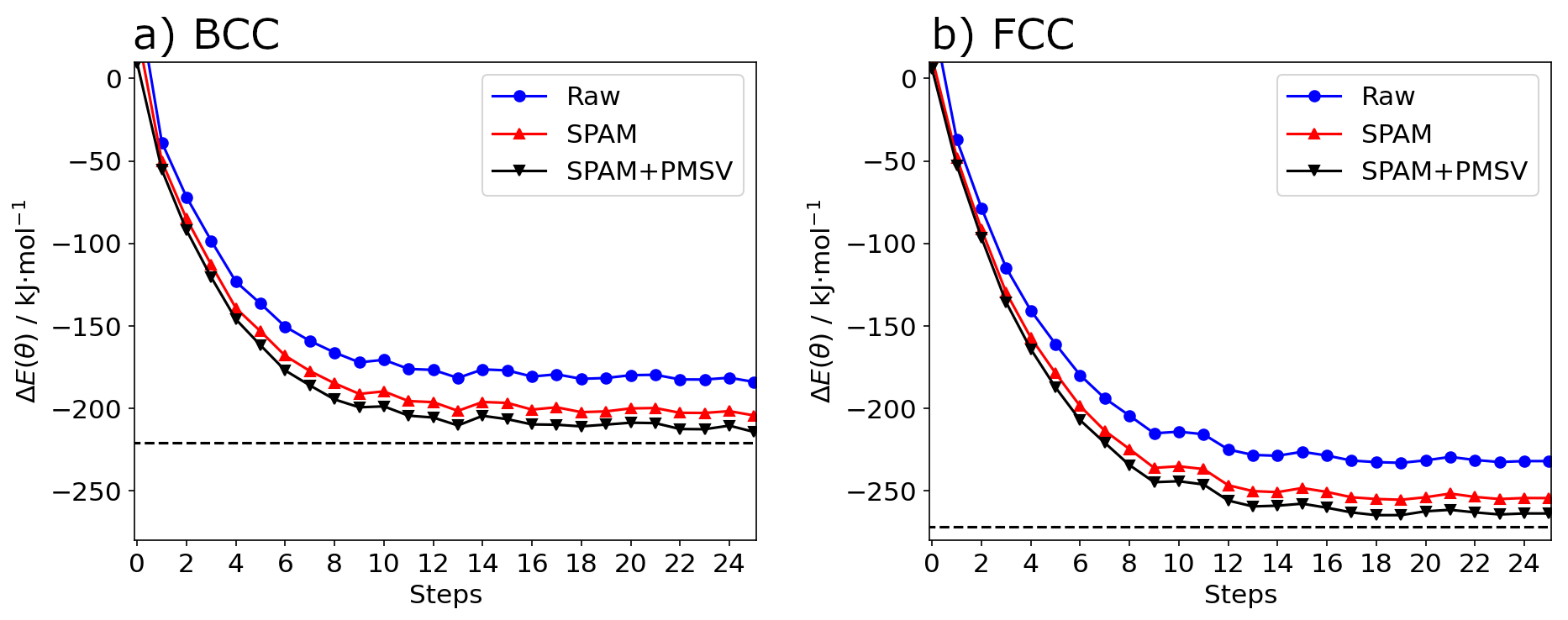}
    \caption{
        Evolution of $\Delta E$ with respect to the number of steps used by the SDG optimization process.
        $(E_{\mathrm{total}}, E_{\mathrm{HF}}^{\circ})
        % =(-220.5~\mathrm{kJ/mol}, -322636.8~\mathrm{kJ/mol})$
         =(-3.228573\times 10^{5}~\mathrm{kJ/mol}, -3.226368\times 10^{5}~\mathrm{kJ/mol})$
        for BCC and
        $(E_{\mathrm{total}}, E_{\mathrm{HF}}^{\circ})
        %=(-271.5~\mathrm{kJ/mol}, -322473.5~\mathrm{kJ/mol})$
        =(-3.227450\times 10^{5}~\mathrm{kJ/mol}, -3.224735\times 10^{5}~\mathrm{kJ/mol})$
        according to the state-vector simulations.
        Black points show hardware results obtained with the \texttt{ibmq\_casablanca} device. The black line is included as a visual guide to show the
        progress of the correlation energy estimate along the variational optimization.
        The dashed line is the correlation energy $E_{\mathrm{corr}}$ for the given model Hamiltonian.
        Error mitigation based on
        SPAM and PMSV
        are applied at the same time.
        Raw and SPAM-corrected $\Delta E$ are given for each optimization step
        to show the effect of noise mitigation.
    }
    \label{fig:exp_qng}
\end{figure*}

Similarly, the SDG optimizer can also accurately locate the ground state
obtained in the reference state-vector simulation.
The results of the progress of $\Delta E(\theta)$
along the variational optimization are shown in
Fig. \ref{fig:exp_qng}(a) for BCC and Fig. \ref{fig:exp_qng}(b) for FCC.
The resulting energies are also in good agreement with the exact results from
state-vector simulations.
In contrast to Rotosolve, the SDG method uses
energy derivatives with respect to the circuit parameter,
which is also evaluated by using the quantum computer.
\textcolor{black}{The resulting parameter and energy are represented as
$\Delta E(-0.5196) = -214 \pm 3$ kJ/mol for BCC and
$\Delta E(-0.6060) = -264 \pm 2$ kJ/mol for FCC}.
The resulting energy of SDG agrees with that of Rotosolve for each of BCC and FCC.
The parameters look different, but it is likely due to the small gradient around the optimal point, which is implied by the slow improvement of the energy in Fig. \ref{fig:exp_qng}.
% It is possible that the gradient is influenced by noise-induced error and
% prevents us from determining the energy as accurately as in the Rotosolve case. 
The SDG optimizer takes more steps in this 1-parameter case,
but it may become more efficient if we take more excitation operators
(circuit parameters) into account in the ansatz.
\textcolor{black}{As in the case of the hydrogen chain, the percentage of discarded shots due to the PMSV correction is low, of the order of $\sim 5\%$ for this system.}

%There is a room for improvement to implement the PMSV. It is theoretically straightforward to apply it
%to the gradient calculation.
%This extension will be considered we we address more parameters with
%deeper quantum circuits.
%Implementation of further noise mitigation for the QNG optimizer is now work
%in progress.

% Note that the QNG method in this case is equivalent to the gradient descent method,
% because the metric tensor is an identity matrix due to the presence of only one
% parameter.
In this particular case, additional excitation operators to the ansatz
would not have a significant effect on the energy ($\sim$5\% of correlation improvement
compared to the first one).
Larger hardware experiments with
more qubits and more excitation operators
(i.e., wider and deeper quantum circuits)
with further noise mitigation techniques including PMSV
are the subject of future work, and will be reported elsewhere.

%=====================================================f=========================
\section{Conclusions and outlook}
\label{sec:conclusions}
%==============================================================================

We have performed quantum hardware calculations for
the solid-state model systems with periodic boundary conditions (PBC).
We employed a distorted hydrogen chain and iron crystals as the
targets, which are both systematically simplified to set a starting
point for simulations on current quantum hardware
with 2 qubit 1 parameter ans\"{a}tze.
We have applied the TransQSE method
for the hydrogen chain
and the PBC-adapted VQE method for the iron crystals.

To make the most of NISQ devices, we have applied
noise mitigation techniques to improve the agreement between the experimental energy estimate
and reference state-vector simulation values.
We have applied 
the state preparation and measurement (SPAM) correction for all the
variational optimizations on hardware.
In all the cases the energy estimates were improved.
In addition to SPAM correction,
we applied the novel PMSV noise mitigation technique,
in which shot counts are post-selected so that the
intrinsic symmetries of the system ($Z_{2}$ and $U_{1}$) are not violated.
SPAM and PMSV are simultaneously used
to realize $\sim 5$ kJ/mol of agreement in the calculations of
simplified iron crystal models.
Although these results are for very simple model systems,
we believe that these results set an important starting point
for systematic improvement of quantum chemical calculations on quantum
computers by rolling back the simplification procedure presented in this paper. Once quantum computers improve, larger basis sets and k-point grids can be considered, and much more accurate estimates of the total energy for these systems can be obtained.

% For an H$_{2}$ lattice, we took
% an H$_{2}$ chain for the target of
% TransQSE experiments,
%into our proprietary quantum chemistry package \texttt{InQuanto},
%and applied it for a simplified iron system
%to evaluate the energy difference between
%body-centered cubic (BCC) and face-centered cubic (FCC) iron crystals.
% The PBC-adapted VQE was properly implemented, and tested with a variety
% of simple systems including H$_{2}$, He, and LiH lattices.
% Technically, the PBC-adapted VQE required
% several modifications for the molecular VQE, which had been already implemented.
% The Hamiltonian and ans\"{a}tz becomes k-point dependent.
% We have implemented k-point dependent Hamiltonian
% by using the integrals obtained by the
% k-point dependent SCF calculation evaluated with \texttt{PySCF},
% we have implemented the k-point dependent
% ans\"{a}tz.
% The amplitudes are generally complex numbers
% because of the nature of Fourier transform \cite{manrique2020momentum}.

In addition to the hardware improvement itself,
several techniques would be useful to extend the range of periodic systems that we may
simulate with quantum computers in the near term.
One of the promising approaches to handle a large system is
using quantum embedding techniques like
density matrix embedding theory (DMET) \cite{knizia2012density} or
Dynamical Mean-Field Theory \cite{georges1996dynamical,rungger2019dynamical}.
For example, the complete unit cell energy in DMET is calculated
as the sum of fragment energies.
The interaction between each fragment and the rest of the unit cell
(referred to as bath) is calculated self-consistently.
The flexibility of the choice of the size of the
fragment allows us to reduce the quantum resource
requirements to address bigger unit cells step by step.
Similar gains can be obtained by using other embedding techniques.

We highlight that consideration of the translational nature of periodic systems, both in direct and reciprocal space, has allowed us to apply simplifications that mitigate the additional resources needed to treat our extended systems on a quantum computer. This is a significant advantage over the consideration of molecular systems with similar sizes in terms of basis sets and numbers of qubits required to perform a simulation. More generally, exploitation of quantum resource savings derived from translational symmetry should always be taken into account when one deals with PBC systems.

Now that all the basic tools to perform PBC-adapted VQE calculations have been tested on hardware,
we expect that larger numbers of basis functions and k-points with wider active spaces will be possible as hardware improves.
Among others,
bigger hardware calculations with more complex
ans\"{a}tze and a variety of noise mitigation techniques are underway in order to go
one step ahead, which will be reported elsewhere.

\section*{Acknowledgments}

We appreciate useful discussions with Koji Hirano from Nippon Steel Corporation. 
From Cambridge Quantum Computing Ltd., we are grateful to Will Simmons
for the discussion on PMSV noise mitigation,
Silas Dikes for useful discussion on SPAM
implementation of \texttt{tket},
and Cono Di Paola and Michal Krompiec for useful comments and help with figures.
The numerical simulations in this work were performed on Microsoft Azure Virtual Machines provided by the program Microsoft for Startups.
% https://www.microsoft.com/en-us/quantum/quantum-network
We acknowledge the use of IBM Quantum services for this work. The views expressed are those of the authors, and do not reflect the official policy or position of IBM or the IBM Quantum team.

\bibliography{main}

% \clearpage

\appendix

%=====================================================f=========================
\section{Partition Measurement Symmetry Verification algorithm details}
\label{appendix:sec:pmsv}
%=====================================================f=========================

Symmetry verification in the NISQ era aims to post-select on quantum measurement results based on a 
measurable property of a quantum circuit during runtime of the computation. The verification procedure reads out a conserved 
property during the quantum calculation
with respect to the projection $\hat{P}$ expressed as  
\begin{equation}
    \hat{P} = \frac{1}{2} \left[ \hat{I} + (-1)^x \hat{S} \right], 
\end{equation}
where $\hat{I}$ is the identity operation, and $\hat{S}$ is a symmetry operation with the parity represented by $(-1)^{x}$, with $x$ being an integer.
Techniques in literature include symmetry verification via mid-circuit measurements or ancilla-controlled 
operations plus readouts \cite{Bonet_Monroig_2018}. The inherent problem with these techniques is the introduction of additional two-qubit gate operations. 
Another difficulty is identifying what symmetries a system possesses 
and how to represent them as operations on many-qubit systems, which may be a non-trivial process. For applications in quantum chemistry, point-group symmetries and space-group symmetries 
can be used to define the symmetries for systems of interest. These can be represented as operations on many-qubit systems by 
following prescribed routines available in literature \cite{setia2020reducing}.

We introduce a technique that can symmetry-verify a quantum circuit without imposing the extra two-qubit gate penalty. In the NISQ era, 
this is a necessity for any simulation that aims to preserve particular properties. An example would be the use of quantum simulations to find 
ground states for molecular and material science problems. Our symmetry verification routine makes use of a measurement reduction
technique implemented in \texttt{tket}. This technique finds partitions of commuting Pauli strings and maps them to
measurement circuits \cite{cowtan2020generic}.
This method is referred to as Partition-Measurement Symmetry Verification (PMSV).

In PMSV, one partition of commuting Pauli-strings corresponds to one measurement circuit. To make a measurement 
circuit ``symmetry-verifiable'', we check if the partition of commuting Pauli strings commutes with every Pauli symmetry that a 
chemical system exhibits. If this check is true, these Pauli-symmetries can be measured by the measurement circuit as well as the Pauli strings of the Hamiltonian. 
When we process the quantum measurement result, we measure the parity, $x$, of our Pauli-symmetry first. We use this check as a constraint 
to keep or discard a particular measurement result, before post-processing the expectation value of the other Pauli strings.
We observe 
that our symmetry verification scheme can be performed  for any quantum simulation via the application of a small classical processing step, but without extra quantum requirements.

The algorithm to prepare a set of quantum circuit and result mapping
is given in Algorithm \ref{algorithm1}, whereas
that to post-select the measurement result is shown in Algorithm \ref{algorithm2}.
% \begin{equation}
%     \hat{S} = (-1)^x \bigotimes_{i} \hat{Z}_i
% \end{equation}

\begin{algorithm}
    \caption{Prepare Symmetry Verifiable Measurement Circuits}
    \label{algorithm1}
    \begin{algorithmic}[1] % The number tells where the line numbering should start
        \State \textbf{Result:} \{Measurable Quantum Circuits, Measurement Result Map\}
        \State Initialize collection of Pauli strings to measure, \{$\hat{P}_{0},\dots,\hat{P}_{N-1}$\}
        \State Initialize collection of Pauli strings based on commuting sets, \{$SET_{0},\dots,SET_{M-1}$\}
        \State Initialize Pauli symmetries, \{$\hat{S}_{0},\dots,\hat{S}_{K-1}$\}
        \For{$SET$ in \{$SET_{0},\dots,SET_{M-1}$\}}
            \For{$\hat{S}_{j}$ in \{$\hat{S}_{0},\dots,\hat{S}_{K-1}$\}}
                % \State BoolianArray[$N$] $symmetry\_commutes$ $\gets$ \{$False,\dots,False$\}
                \State $all\_commutes \gets True$
                \For{$\hat{P}_{i}$ in \{$\hat{P}_{0},\dots,\hat{P}_{N-1}$\}}
                    \If{$[\hat{S}_{j}, \hat{P}_{i}] \ne 0$}
                        \State $all\_commutes \gets False$
                        \State \textbf{break}
                    \EndIf
                \EndFor
                \If {$all\_commutes$}
                    \State Insert $\hat{S}_{j}$ in $SET$
                \EndIf
            \EndFor
            \State $Circuit, Map \gets partition\_to\_measurement(SET)$
        \EndFor
    \end{algorithmic}
\end{algorithm}

\begin{algorithm}
    \caption{Post-Select Measurement Result}
    \label{algorithm2}
    \begin{algorithmic}[1] % The number tells where the line numbering should start
        \State \textbf{Result:} \{Symmetry-Verified Quantum Measurement result\}
        \State Initialize map from measurement result to Pauli-symmetries result, $map\_list$
        \State Initialize quantum measurement result, $result\_list$
        \State Initialize expected parity of Pauli Symmetries, $targets$
        \For{$result$ in $result\_list$}
            \State $pauli\_parities \gets \{\}$ (empty list)
            \For{$map$ in $map\_list$}
                \State Insert $map(result)$ in $pauli\_parities$
            \EndFor
            \If {$pauli\_parities \ne targets$}
                \State discard $result$ from $result\_list$
            \EndIf
        \EndFor
    \end{algorithmic}
\end{algorithm}

We demonstrate how PMSV works by comparing the probabilities of the measurement outcome from the noisy simulation with different error mitigation techniques.
Single point energy calculation with the preliminary optimized
parameter (by state-vector simulation) is considered for convenience.
Noise model is taken from the IBM device identified with
\texttt{ibmq\_casablanca}.
Two simplified systems (the hydrogen chain and iron BCC) used in the main text are taken as examples.
In both cases we have two circuits to be measured for both evaluating expectation value
and symmetry-verification.

As shown in Tab. \ref{tab:pmsv_h2_chain},
the result without error mitigation (Raw) includes nonzero probabilities
for ``01'' and ``10'' of Circuit (1), and ``01'' and ``11'' of Circuit (2).
These correspond to symmetry-violating states that must be caused by the noise.
When PMSV is applied, only the symmetry-verified states are post-selected,
but the proportion of the probabilities for each circuit is not in good agreement
with the corresponding noiseless results.
SPAM correction improves the probability both in the proportion of the probabilities and
the reduction of the symmetry-violating states.
Therefore no significant improvement was made by PMSV followed by SPAM
in this particular case.
\begin{table*}[]
\caption{
    Probabilities of the measurement outcome from the noisy simulation of the simplified hydrogen chain system
    with the preliminary optimized parameter $\theta=-0.09283$.
    See Sec. \ref{sec:h2_transqse} for the computational details.
    Noise model is taken from the IBM device identified with \texttt{ibmq\_casablanca}.
    ``--'' indicates a value of zero.
    Noiseless simulation result is given for reference.
    \label{tab:pmsv_h2_chain}
    }
\begin{tabular}{l|cccc|cccc|c}
\hline
    & \multicolumn{4}{c|}{Circuit (1)} & \multicolumn{4}{c|}{Circuit (2)} \\ \hline
            & 00        & 01        & 10        & 11        & 00        & 01        & 10        & 11    & $\Delta E$ / kJ$\cdot$mol$^{-1}$\\ \hline
Raw         & 0.9657    & 0.0105    & 0.0143    & 0.0095    & 0.4290    & 0.0009    & 0.5543    & 0.0077    & $24.4$  \\
PMSV        & 0.9892    & --        & --        & 0.0108    & 0.4363    & --        & 0.5637    & --        & $-7.3$ \\
SPAM        & 0.9903    & 0.0001    & 0.0001    & 0.0095    & 0.4241    & 0.0002    & 0.5755    & 0.0001    & $-26.0$\\
SPAM+PMSV   & 0.9905    &  --       & --        & 0.0095    & 0.4243    & --        & 0.5757    & --        & $-26.3$ \\    \hline
Noiseless   & 0.9910    &  --       & --        & 0.0090    & 0.4094    & --        & 0.5906    & --        & $-41.7$\\     \hline
\end{tabular}
\end{table*}

The situation is different in the iron simulations. Tab. \ref{tab:pmsv_iron_bcc},
shows that SPAM does not reduce the number of symmetry-violating states as much as in the hydrogen chain case,
and we find that the SPAM+PMSV combination gives us the lowest energy, whereas all the other tendency
is the same.
We can understand this difference by comparing the magnitude of the parameter $\theta$.
As the parameter $\theta=-0.09283$ of the hydrogen chain case
is smaller than that of the iron case, $\theta=-0.53038$ in magnitude, noise induced by SPAM would be the major source
of error in the former case, so that SPAM correction works well and minimises the room
for PMSV to improve the result.
\begin{table*}[]
\caption{
    Probabilities of the measurement outcome from the noisy simulation of the simplified iron BCC system
    with the preliminary optimized parameter $\theta=-0.53038$.
    Noise model is taken from the IBM device identified with \texttt{ibmq\_casablanca}.
    See Sec. \ref{sec:fesystems_2k} for the computational details.
    ``--'' indicates a value of zero.
    Noiseless simulation result is given for reference.
    \label{tab:pmsv_iron_bcc}
    }
\begin{tabular}{l|cccc|cccc|c}
\hline
    & \multicolumn{4}{c|}{Circuit (1)} & \multicolumn{4}{c|}{Circuit (2)}  \\ \hline
            & 00        & 01        & 10        & 11        & 00        & 01        & 10        & 11        & $\Delta E$ / kJ$\cdot$mol$^{-1}$\\ \hline
Raw         & 0.7203    & 0.0151    & 0.0208    & 0.2437    & 0.0967    & 0.0063    & 0.8883    & 0.0088    & $-187.5$ \\
PMSV        & 0.7472    & --        & --        & 0.2528    & 0.0981    & --        &  0.9019   & --        & $-196.3$     \\
SPAM        & 0.7336    & 0.0036    & 0.0073    & 0.2555    & 0.0713    & 0.0055    & 0.9177    & 0.0055    & $-208.8$\\
SPAM+PMSV   & 0.7417    &  --       & --        &  0.2583   & 0.0721    & --        & 0.9279    & --        & $-213.2$ \\ \hline
Noiseless   & 0.7447    &  --       & --        &  0.2553   & 0.0621    & --        & 0.9369    & --        & $-221.0$ \\ \hline
\end{tabular}
\end{table*}

%=====================================================f=========================
\section{DFT/PBE calculations to estimate the required k-points}
\label{appendix:sec:dft}
%=====================================================f=========================

To support that the [4~4~4] k-point mesh for UHF/CCSD/LANL2DZ iron calculations
would be sufficiently dense,
instead of performing larger UHF/CCSD calculation (which is found to be unfeasible),
we perform an exploration with unrestricted DFT
(U-DFT) using Perdew–Burke-Ernzerhof (PBE) functional \cite{perdew1996generalized}
to check if the potential curves are stationary as k-point number and/or basis set size increase.
We compare the BCC--FCC energy difference and the lattice constants
obtained with various k-point meshes, namely,
$[L_{m}~L_{m}~L_{m}]$ Monkhorst-Pack meshes with $L_{m} = \{1, 2, 3, 4, 5\}$.

\begin{figure}
    \centering
    \includegraphics[width=0.84\hsize]{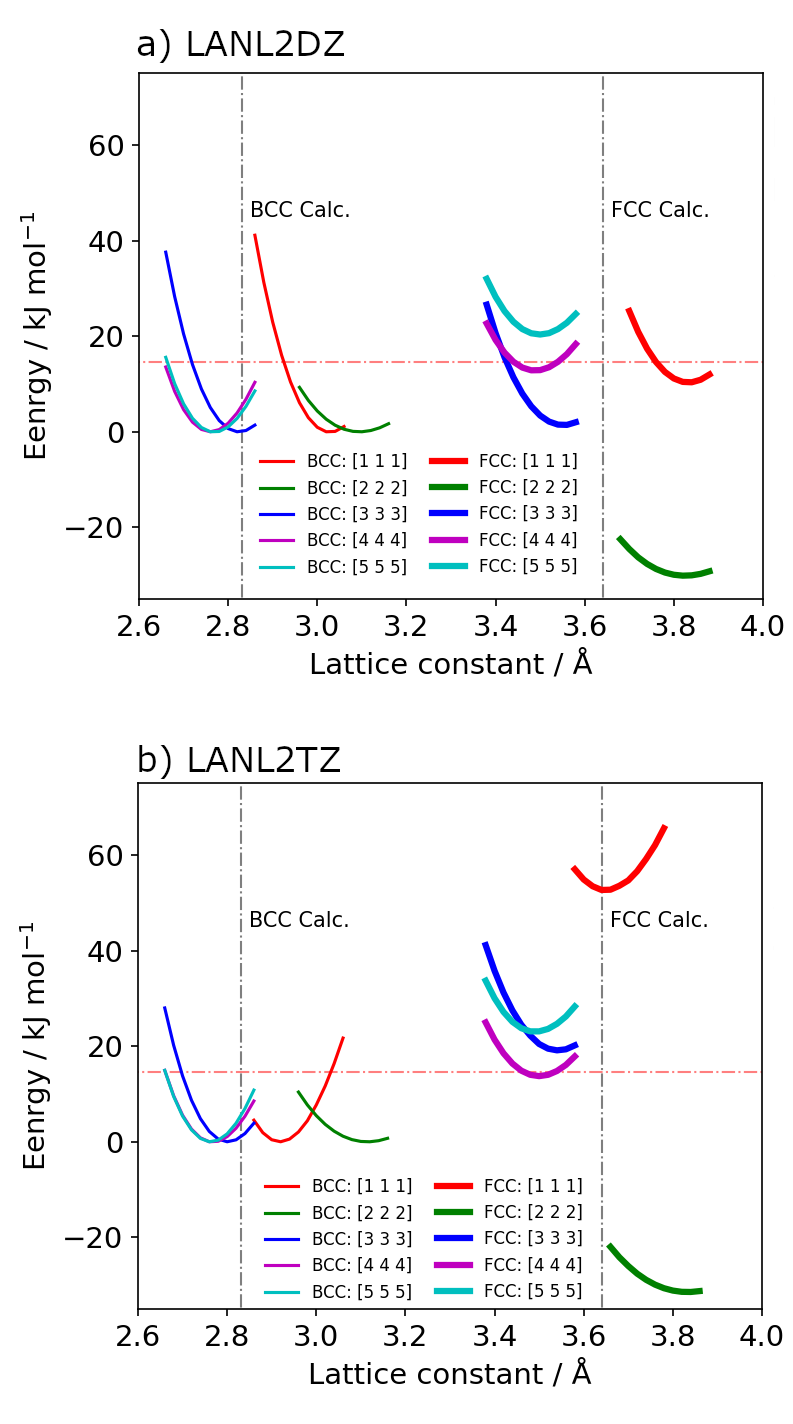}
    \caption{
        Potential energy curves 
        as functions of 
        lattice constants for various k-point mesh
        indicated by $[L_{m}~L_{m}~L_{m}]$
        obtained with the U-DFT/PBE level of theory with (a) LANL2DZ
        and (b) LANL2TZ basis sets.
        The optimized lattice constants
        from the accurate
        DFT calculations from the literature \cite{nguyen2018first}
        are indicated by vertical lines,
        while the BCC--FCC energy difference is 
        shown in the
        horizontal line on each panel.
        Energy baseline is set so that the energy of optimal lattice constant of BCC
        is equal to zero for each $L_{m}$.
    }
    \label{fig:pes_dft}
\end{figure}

The potential energy curves obtained with the
double-zeta basis (LANL2DZ) (Figure \ref{fig:pes_dft}(a))
and
triple-zeta basis (LANL2TZ) (Figure \ref{fig:pes_dft}(b))
indicate that a
[4~4~4] k-point mesh with the LANL2DZ basis set would be sufficient for
stationary estimation of the BCC--FCC energy difference.

The [1~1~1] k-point ($\Gamma$-point)
with the LANL2DZ basis gives a reasonable result in the BCC--FCC energy difference,
but it changes significantly with the LANL2TZ basis.
The lattice constants are overestimated up to 0.2 {\AA} in both BCC
and FCC cases.
The [2~2~2] k-point mesh calculations commonly return
the opposite energy difference
in DZ and TZ basis sets.
The lattice constants are similarly overestimated.
The [3~3~3] k-point mesh still returns the opposite energy difference
with DZ basis, but with TZ basis it is in good agreement with the value from the
literature \cite{nguyen2018first}.
The lattice constants turn to be underestimated up to 0.1 {\AA}.
The [4~4~4] and [5~5~5] k-point mesh calculations
commonly return the accurate energy differences both with DZ and TZ basis.
As a consequence, 
we have focused on the [4~4~4] k-point mesh with LANL2DZ basis set
in the iron calculations in the main text.

\end{document}